\documentclass[twocolumn,longauthor]{aastex631}
\usepackage[utf8]{inputenc}
\defcitealias{K16}{I}
\defcitealias{K18}{II}
\defcitealias{K22}{IIa}

\shorttitle{Cosmic ray heating of mixed grains}
\shortauthors{Kalv\=ans \& Kalnin}
\graphicspath{{./}{figures/}}

\begin{document}

\title{Temperature Spectra of Interstellar Dust Grains Heated by Cosmic Rays.\\
III. Mixed Composition Grains\footnote{Received 2022 February 11; revised 2022 September 6; accepted 2022 September 13}}

\correspondingauthor{Juris Kalv\=ans}
\email{juris.kalvans@venta.lv}

\author[0000-0002-2962-7064]{Juris Kalv\=ans}
\affil{Engineering Research Institute ``Ventspils International Radio Astronomy Center'' of Ventspils University of Applied Sciences,\\
In$\check{z}$enieru~101, Ventspils, LV-3601, Latvia}

\author{Juris Roberts Kalnin}
\affil{Engineering Research Institute ``Ventspils International Radio Astronomy Center'' of Ventspils University of Applied Sciences,\\
In$\check{z}$enieru~101, Ventspils, LV-3601, Latvia}

\begin{abstract}
Icy grains in the interstellar medium and star-formation regions consist of a variety of materials. Such composite grains interact differently with cosmic-ray (CR) particles compared to simple single-material grains. We aim to calculate the spectra of energies and temperatures of mixed-composition grains undergoing whole-grain heating by CRs. The grains were assumed to consist of a mixture of carbon and olivine, covered by ices consisting of carbon oxides and water. The energy and temperature spectra for grains with radii 0.05; 0.1, and 0.2 microns impacted by CRs were calculated for eight values of column density, relevant to molecular clouds and star-forming cores. The approach takes into account changes in ice thickness and composition with increasing column density. These detailed data for CR interaction with interstellar grains are intended for applications in astrochemical models. The main finding is that the a more accurate approach on grain heat capacity and other factors prevent a frequent heating of 0.1 micron or larger icy grains to high temperatures.
\end{abstract}


\section{Introduction} \label{intrd}

Cosmic-ray (CR) induced whole-grain heating of grains is a process that is able to significantly affect the balance between gaseous and solid ice phases in dense molecular clouds and star-forming regions \citep{Hasegawa93}. In particular, an efficient sublimation and the resulting retention of the volatile CO molecule in the gas-phase influences the whole chemical evolution of a dense cloud core. Such CR-induced desorption (CRD) has since been adopted in most astrochemical models in use. Therefore, an accurate description of this process is of high importance in theoretical astrochemistry. Whole-grain heating affects also the rates of diffusion and chemical reactions in the icy mantles of interstellar grains.

In the recent years, the underlying physics for the CR-induced grain heating process has received renewed interest. \citet{Ivlev15} investigated the role of CRs as the triggering factor for chemical explosions in the mantles of icy grains. Detailed temperature and energy spectra of grains undergoing whole-grain heating by CRs were published by \citet[][hereinafter Paper~I]{K16}, \citet[][Paper~II]{K18}, and its data correction \citet[][Paper~IIa]{K22}. The studies by \citet{Iqbal18}, \citet{Zhao18}, \citet{Sipila20}, and \citet{Silsbee21} are motivated to address in detail the heating and cooling of grains by their aim to investigate the chemical differentiation of ices across grain sizes. In \citet{Kalvans19} and \citet{Kalvans21} we employed the data of Paper~\citetalias{K18} for quantifying whole-grain heating in astrochemical models assuming a single size of grains. A similar work with a different approach is that of \citet{Sipila21}. A number of theoretical studies also consider the specifics of CR interaction with icy interstellar grains. \citet{Reboussin14,K15aa,Ivlev15} and \citet{Shingledecker17} investigate the chemical processing of ices caused by CRs. While these studies cannot be compared directly with our work, some data overlaps. For example, the energy loss function of protons in water ice from \citet{Shingledecker20} agrees well with our data, despite the different methods employed.

The above studies have demonstrated the need of reliable energy and temperature spectra of CR-heated interstellar grains. While Paper~\citetalias{K18} provide such elaborated data, it has several deficiencies. First, it considers the composition of the grain core and its icy mantle in a simplified manner, by assuming only a single component (olivine) for the refractory grain core. The choice of grain materials affects the energy loss function $L_{\rm grain}$ of the impacting CR particle and the heat capacity $C$ of the grain, which affect the temperature spectra of CR-heated grains and thus the efficiency of CRD. Second, in a recent work \citep{Kalvans20}, we found that $C$ of a Debye solid, used in Paper~\citetalias{K18}, is not suited for the temperatures involved in whole-grain heating by CRs and that less-general and material-specific $C$ data are more reliable.

The aim of this paper is to provide accurate data on CR-induced interstellar whole-grain heating for calculating the efficiency of CRD. We consider grains of mixed composition, which affects their $L_{\rm grain}$ and $C$. In addition, the composition of the icy mantle changes with different interstellar extinctions $A_V$. While this may have a limited effect on $L_{\rm grain}$, the heat capacities of the primary ice components H$_2$O, CO, and CO$_2$ are quite different, which affects the grain's initial peak temperature reached immediately after the CR hit.

Corresponding to the above aim, the tasks of this study are choosing feasible chemical and physical properties of icy grains in dense cores, quantifying their interaction with CRs, and presenting the essence of the results in an understandable manner. The results will permit more accurate modeling of CRD and other CR-induced processes on interstellar grains by providing the energy received and temperature reached by the grains with ice mantle properties relevant to the hydrogen column density $N_{\rm H}$ and interstellar extinction $A_V$ in their environment.

\section{Methods} \label{mthd}

Cosmic-ray particles of various chemical elements, each with an initial energy spectrum, travel large distances through the tenuous interstellar medium (ISM) consisting of H, He, and traces of other elements. The CR particles interact with the ISM, which shifts higher-energy particles to lower energies and absorbs the lowest-energy particles. We are interested in CRs that impact a grain after traveling a certain amount of $N_{\rm H}$. Observations show that the existence and properties of the ice layer on grains depend on the $A_V$ \citep[][see also Table~A1 of \citeauthor{K18mn} \citeyear{K18mn} and references therein]{Whittet01}, which means that the icy mantle will have a composition and thickness that depends on $N_{\rm H}$.

The CR particles may impact a single icy grain at different angles, traversing tracks of various lengths. Some of the CR tracks will affect only the ice layer not touching the refractory grain core at all. As a CR particle traverses, it interacts with grain material, losing a certain amount of energy. While some of the energy escapes in the form of fast electrons and ejected ice molecules, the rest eventually thermalizes and the grain attains an uniform temperature within $\approx10^{-9}$\,s \citep{Bringa04}. For most cases, it can be assumed that no cooling occurs up to this point because even at $>$120\,K significant evaporation of volatiles commonly encountered in interstellar ices require longer timescales \citep{Kalvans20}. The thermal energy of the grain $E_{\rm grain}$ is the basic outcome of our calculation. With the help of $C$, it is transformed into $T_{\rm CR}$, which is the initial or peak temperature of CR-heated grains for astrochemical models considering CRD and can be considered as a more applicable representation of $E_{\rm grain}$. Because different parts of the grain are bombarded by CRs of various elements and energies, the obtained $E_{\rm grain}$ and $T_{\rm CR}$ can be most accurately presented in the form of energy and temperature spectra, respectively.

\subsection{Grain model} \label{mdl}

\begin{figure}[ht!]
\vspace{2cm}
\plotone{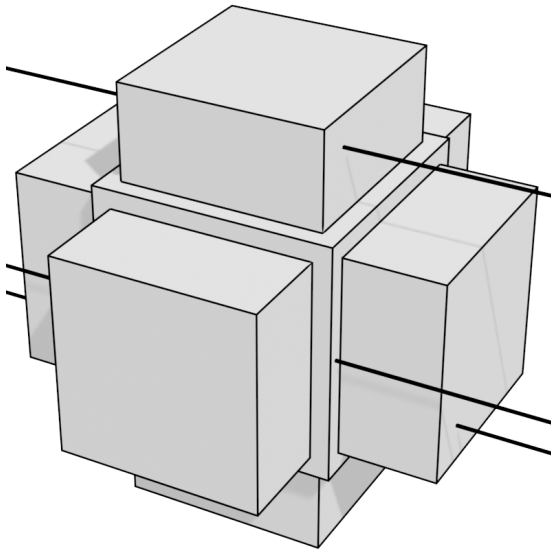}
\caption{Scale model of the core of the interstellar grain. The thick lines show three types of allowed CR tracks through the grain. Track types along the other two symmetry axis are also possible but are not shown for clarity.}
\label{fig-mdl}
\end{figure}
Similarly to our previous studies, three grain radii $a$ were considered -- 0.05, 0.1, and 0.2\,$\mu$m. This means that the data are optimized for application in common, general astrochemical models that consider a single grain size of about 0.1\,$\mu$m. The data are not meant for specialized simulations investigating chemistry with different grain sizes, such as those mentioned in Section~\ref{intrd}, or other advanced models considering a grain size distribution \citep[e.g.][]{Pauly18}. Chemical modeling with different grain sizes results in a different ice composition and thickness for each grain size \citep{Iqbal18,Zhao18}. This effect is yet to be quantified for the variety of $A_V$ and density combinations encountered in evolving dense cores. Similar ice thickness for all grain sizes can be expected in a case when desorption does little to hamper ice accretion \citep{Acharyya11,Pauly16}. Moreover, there is suspicion that small grains are depleted via grain-grain collisions in molecular clouds \citep{Silsbee20}. Therefore we consider describing CR-induced grain heating for relatively large grains with similar ice thickness at similar $A_V$ as a safe choice.

Observations and models indicate that at increasing interstellar extinctions $A_V$, molecules locked in icy mantles covering the grains become more abundant and with a higher proportion of ``hypervolatile'' icy species, such as CO, N$_2$, O$_2$, and CH$_4$. It is not possible to calculate $T_{\rm CR}$ and $E_{\rm grain}$ spectra for all possible combinations of grain mantle thickness and ice composition at different $A_V$ (and thus, column densities $N_{\rm H}$). We overcome this problem by adopting feasible average, characteristic ice composition for several $A_V$ values. The derivation of ice properties is described in Section~\ref{chm}.

The calculation of the interaction between mixed-composition grains and CRs is a non-trivial undertaking when the full spectra of multiple CR species is considered. To our knowledge, all previous studies, starting with that of \citet[][except the series starting with Paper~\citetalias{K16}]{Leger85}, have considered only a single path for CRs hitting the grain -- that through the center of a spherical homogeneous grain. The length of this path is equal to grain diameter, while its area is the maximum cross-section of the spherical grain. Such assumptions mean that, in practice, a cylinder with radius $a$ and length $2a$ is considered. The energy left by CRs in such a cylindrical grain is higher by a factor of 1.5 (the ratio of volumes) compared to that of a spherical grain. Moreover, this approach ignores CR paths passing through ice, without touching grain nucleus. Such paths add lower-temperature grain heating events. These errors can be avoided by considering more than one CR path through the grain. A simple yet efficient way to do this is to consider a grain model consisting of several cuboids. It is possible to adjust the size of the cuboids so that the key parameters -- CR track length, determining the energy acquired by the grain, and grain cross section, determining collision frequency, are very close to those of a sphere. The track length and cross-sections of the cuboids produce the volume of the grain. In Paper~\citetalias{K16}, the grain model considered two CR tracks through the grain core and one or two tracks through the ice layer, depending on ice thickness. This approach decreased the error to less than 10\,\% in CR track length, overall grain cross-section, and volume, compared to a spherical grain.

Continuing the approach of Paper~\citetalias{K16}, we consider several tracks for CR particles passing through cuboid grain models that have geometrical properties representative for spherical bodies. The model was constructed by attaching cuboids to the faces of a central cube, and allowing the CR particles to enter the grain model only at right angles. This time, to make the calculations more accurate, we consider three CR tracks through the grain core. Figure~\ref{fig-mdl} depicts the basic grain core model to scale. The cross-section of the central cuboid with length 2$a$ is 0.339 relative to that of a circle. This means that the frequency of CR hits that can heat the grain to maximum temperatures is reduced by about a factor of 3, compared to the above-described approach considering a grain cylinder.

For icy grains, cuboids consisting of ice were added to the base model applying the five principles laid out in Section~2.2 of Paper~\citetalias{K16}. Shortly, this means that the surface area, maximum CR track length through grain core and ice, and grain volume are similar to the corresponding values of spherical grains covered with icy mantles of uniform thickness. This required also including a CR track that passes through only icy matter, without touching the refractory grain core (cf. Figure~3 of Paper~\citetalias{K18}). To maintain the deviation of the above-listed grain model geometrical parameters within 3\,\% from those of spherical grains, we found it necessary to add two such icy CR tracks for the smaller 0.05\,$\mu$m grains. Table~\ref{tab-nh} shows the number of CR tracks considered for each grain size at each $A_V$ value. {Considering more than a few CR tracks further complicates calculations, while giving increasingly lower benefits in data accuracy and thus is not cost-effective.}

Each cuboid has its own specific cross section $S$ and track length $s$ for calculating the energy left by a variety of CR species with a range of energies (see Section~\ref{crsp}). Even the simplest bare grain has close to 30,000 entries for CR-grain interactions. Impacts of CRs resulting in similar grain energies $E_{\rm grain}$ and, thus, heating temperatures $T_{\rm CR}$ (Section~\ref{grloss}) were grouped together with their frequencies summed up, producing $E_{\rm grain}$ and $T_{\rm CR}$ spectra (Section~\ref{rslt}). The properties of the different CR tracks are described in detail Appendix~\ref{app-a}.

\subsection{Grain composition}
\label{chm}

\begin{table*}
\centering
\caption{Grain parameters for calculating the $T_{\rm CR}$ and $E_{\rm grain}$ spectra for grains hit by CRs. Ice composition and thickness data adopted from \citet{Kalvans21} and were assumed equal for all three grain sizes.}
\label{tab-nh}
\begin{tabular}{lccccccccccc}
\tablewidth{0pt}
\hline
\hline
 &  &  &  & \multicolumn{2}{c}{Grain Core} & \multicolumn{3}{c}{Icy Mantle} & \multicolumn{3}{c}{No. of CR tracks}\tablenotemark{a} \\
No. & $A_V$\tablenotemark{a} (mag) & $N_{\rm H}$ (cm$^{-2}$) & $B$\tablenotemark{b} ($\mu$m) & \% Carbon & \% Silicate & \% H$_2$O & \% CO$_2$ & \% CO  & 0.05\,$\mu$m & 0.1\,$\mu$m & 0.2\,$\mu$m \\
\hline
1 & 1.5 & $3.30\times10^{21}$ & 0 & 40 & 60 & \nodata & \nodata & \nodata & 3 & 3 & 3 \\
2 & 3.0 & $6.60\times10^{21}$ & 0.010 & 40 & 60 & 78 & 17 & 5 & 4 & 4 & 4 \\
3 & 6.0 & $1.32\times10^{22}$ & 0.020 & 40 & 60 & 72 & 14 & 14 & 5 & 4 & 4 \\
4 & 9.0 & $1.98\times10^{22}$ & 0.025 & 40 & 60 & 60 & 17 & 23 & 5 & 4 & 4 \\
5 & 20 & $4.40\times10^{22}$ & 0.030 & 40 & 60 & 50 & 22 & 28 & 5 & 4 & 4 \\
6 & 40 & $8.80\times10^{22}$ & 0.030 & 40 & 60 & 50 & 22 & 28 & 5 & 4 & 4 \\
7 & 80 & $1.76\times10^{23}$ & 0.030 & 40 & 60 & 50 & 22 & 28 & 5 & 4 & 4 \\
8 & 160 & $3.52\times10^{23}$ & 0.030 & 40 & 60 & 50 & 22 & 28 & 5 & 4 & 4 \\
\hline
\end{tabular}
\tablenotetext{a}{ Assuming $N_{\rm H}/A_V=2.2\times10^{21}$\,cm$^{-2}$}
\tablenotetext{b}{ Ice thickness}
\tablenotetext{c}{ Properties of the CR tracks are listed in Table~\ref{tab-tracks} of Appendix~\ref{app-a}.}
\end{table*}

The dust grain refractory cores are described as composite grains consisting of mixed silicate and carbonaceous materials. A similar approach for describing CR interaction with grains was employed by \citet{Shen04}. Such composite grains are supported by some dust models \citep{Li97,Vaidya07,Valencic15}. Even when the grains in the ISM consist of separate carbonaceous and silicate grain populations, the energy budget of CR impacts is similar to that of composite grains as long as the proportions of the chemical materials are equal, although the maximum energies and temperatures can vary. Observations indicate that the fraction of carbon in interstellar dust is 0.3--0.5 \citep[see Table~1 of][and references therein]{Hocuk17}. Here we assume the approximate average value of 0.40.

\begin{table*}
\centering
\caption{Properties of grain materials.}
\label{tab-gr}
\begin{tabular}{llcl}
\tablewidth{0pt}
\hline
\hline
No. & Material & density (g\,cm$^{-3}$) & Heat Capacity Reference \\
\hline
1 & Olivine & 3.32 & \citet{Draine01} and \citet{Xie18} \\
2 & Carbon & 1.56 & \citet{Draine01} and \citet{Xie18} \\
\hline
3 & H$_2$O ice & 0.90 & \citet{Shulman04} \\
4 & CO$_2$ ice & 1.60 & \citet{Giauque37} \\
5 & CO ice & 1.10 & \citet{Clayton32} \\
\hline
\end{tabular}
\end{table*}
%
\begin{figure}[ht!]
\vspace{2cm}
\hspace{-1cm}
	\includegraphics[width=11cm]{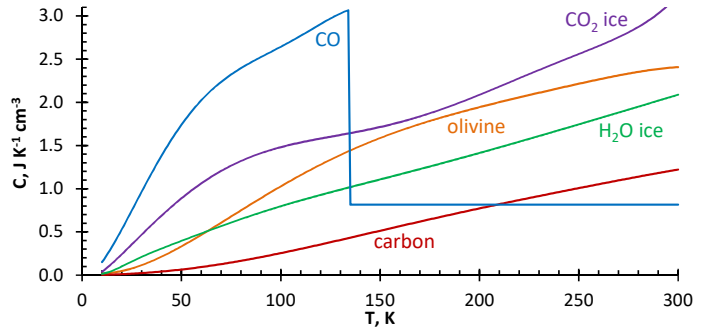}
\caption{Heat capacities $C$ for grain constituents.}
\label{fig-hcap}
\end{figure}

Table~\ref{tab-gr} summarizes the properties of grain materials. Two separate components were considered -- the grain core consisting of a homogeneous mixture the ``astronomical silicate'' olivine MgFeSiO$_4$ and carbon covered by a mantle consisting of a homogeneous mixture consisting of H$_2$O, CO$_2$, and CO ices with proportions listed in Table~\ref{tab-nh}. The density of the mixtures was calculated as the weighed average density of their components.

Heat capacity $C$ was calculated with a different method for each material. For the refractory grain core materials, volumnic heat capacities were applied directly, while for the volatile ices, whose constituent proportions are varied at different $A_V$, $C$ calculation could be conveniently done by first obtaining $C$ per molecule, which was multiplied by the supposed exact number of molecules of a given type on the grain. The $C$ values for CO$_2$ and, in particular, CO ices are substantially higher than those of H$_2$O ice, which is the main reason why these materials were considered in this study. Similarly to density, the $C$ of the mixtures were calculated as the weighed average value. Figure~\ref{fig-hcap} shows that $C$ substantially differs for the ice constituents H$_2$O, CO, and CO$_2$. The high heat capacity of carbon oxides means a lower $T_{\rm CR}$ for the icy grain, compared to our previous calculations, where ice was assumed to have a heat capacity similar to that of pure H$_2$O.

Given that dust grains are subjected to interstellar ionizing radiation, amorphous carbon (aC) is a candidate carbonaceous component \citep{Koike80}. However, no specific approach for calculating $C$ for aC was found in literature. Thus, $C$ was calculated as for graphite, albeit with the density of aC (1.557\,g\,cm$^{-3}$).

In considering the heating and subsequent cooling of icy grains, one has to take into account that CO has its critical point at 134\,K. In our calculations, this $T_{\rm CR}$ value is exceeded by 0.05\,$\mu$m grains 0.01\,$\mu$m thick ice layer. (It is also exceeded for bare 0.05\,$\mu$m grains that do not contain CO.) For CO-rich grains exceeding this temperature, an orderly, layer-by-layer evaporative cooling is likely not possible. This means that some calculation results -- part of the $T_{\rm CR}$ spectra for 0.05\,$\mu$m grains -- can be inaccurate. Nevertheless, the $E_{\rm grain}$ spectrum is accurate because energy deposition occurs on much shorter time-scales \citep{Bringa04} and may still allow investigating processes on such ``overheated'' grains.

We consider water and carbon oxides as the components of interstellar ices. While there have been observed other components, such as methane, methanol, and ammonia, which are important for cloud chemistry, their abundances are lower and less constrained by observations. Therefore, their effects on grain properties are limited and less well quantifiable. Observations of interstellar ices give limited information on their exact composition in the context of their surrounding medium and history. Astrochemical models can be calibrated with the help of the observational data, such as \citet{Boogert11}. Such calibration, coupling ice abundances to $A_V$, allows to interpolate the lacking details of the ``average'' H$_2$O, CO, and CO$_2$ relative abundances at different $A_V$ values with an acceptable degree of credibility.

For obtaining the ice thickness and exact H$_2$O, CO, and CO$_2$ ice relative abundances, the latest version of the model \textsc{Alchemic-Venta} was adapted. A comprehensive description of this model is provided by \citet{Kalvans21}. Shortly, this model simulates the chemistry in a contracting typical low-mass star-forming core with the rate-equation method. A single change was made to the model -- the interstellar photon and CR flux were modified by a factor of 0.2 (instead by 0.1 and 1.0 in that study), which ensured a better agreement of calculated relative abundances of icy species to observational data. The CR-induced processes in this model are $N_H$-dependent. For CRD, the attenuation of CR flux was considered with the data of Paper~\citetalias{K18}.

Ice properties from the model were sampled at $A_V$ values listed in Table~\ref{tab-nh}. The ice thickness of 0.01, 0.02, 0.025, and 0.03 $\mu$m roughly correspond to 30, 60, 75, and 90 ice monolayers. The ice abundances listed in Table~\ref{tab-nh} generally agree with other astrochemical models \citep{Taquet16,Iqbal18,Aikawa20,Wakelam21,O22,Garrod22}, although can be different from models, where the production of some major icy molecule (which often is CO$_2$) has not been solved, or models adjusted to specific tasks \citep[e.g.,][]{Vasyunin17,Sipila20,Rawlings22}. The advantage for using the relative abundances of icy species from \textsc{Alchemic-Venta} is the above-mentioned $A_V$-dependent calibration of ice composition with observational data \citep[for an example, see][]{K18mn}.

The value for $N_{\rm H}/A_V$ was taken to be $1.6\times10^{21}$\,cm$^{-2}$ in the astrochemical model. However, a literature study reveals that this is in the lowest end of possible values. The published $N_{\rm H}/A_V$ values in the ISM range from approximately $1.8\times10^{21}$ \citep{Predehl95} to $5.1\times10^{21}$\,cm$^{-2}$ \citep{Predehl94,Guver09}. Many of these measurements have been made towards X-ray sources \citep[e.g.,][]{Reina73}. Higher $N_{\rm H}$ values, relevant to molecular clouds, tend to have a higher $N_{\rm H}/A_V$ \citep{Valencic15}. Other recent accurate measurements using HI observations sample sight lines with low column densities \citep[$N_{\rm H}<10^{20}$\,cm$^{-2}$;][]{Liszt14,Chen15,Lenz17}, which are less relevant for our dense cloud core. Therefore, as the most reliable data we found the ``gold'' sample of \citet{Zuo21}, who consider $N_{\rm H}>5\times10^{21}$\,cm$^{-2}$ and find $N_{\rm H}/A_V = 2.2\times10^{21}$\,cm$^{-2}$.

Table~\ref{tab-nh} lists ice thickness and proportions between H$_2$O, CO$_2$, and CO along with the respective $N_{\rm H}$ values used for calculating CR spectra in Section~\ref{crsp}. These ice mantle data are most relevant to low-mass interstellar starless and star-forming cores in a quiescent environment. The choice of $A_V$ from 1.5 to 160\,mag corresponds  to molecular clouds and their dense cores, from the onset of water ice formation to prestellar cores. These $A_V$ and their corresponding column density $N_{\rm H}$ values describe attenuation from the edge of the cloud core to its center.

\subsection{Cosmic ray spectra} \label{crsp}

We consider CR particles that proceed through a cloud core in a straight line. In real clouds, the CR paths and attenuation are much more complex. Such aspects were not considered in our calculations. They can result in overall change of CR flux in dense cores, or even just parts of such cores. The effects of magnetic focusing, mirroring, and scattering can have different outcomes \citep{Padovani13,Silsbee18,Ivlev18} and should be accounted for in astrochemical simulations, depending on what type of object is modeled and other considerations. Also, one can take into account that the overall intensity of CRs varies, depending on the location in the Galaxy \citep[e.g.][]{Indriolo15}.

\subsubsection{Initial H spectrum} \label{inisp}

Following an established practice, the energy spectra of all CR elemental components was based on the initial energy spectrum of protons. Our main result is a data set, which employs a basic CR proton spectrum, which we deem most relevant to the ISM. As a default choice, we adapted the ``model high'' spectrum of \citet[][Alexei Ivlev, private communication]{Padovani18}. This spectrum produces a CR intensity for H atoms that is slightly lower by a factor of $\approx1.5$ than the intensity adopted in Papers \citetalias{K16} and \citetalias{K18}. Only the results with the ``model high'' spectrum were discussed, however, for completeness, Appendix~\ref{app-b} shows also the $T_{\rm CR}$ and $E_{\rm grain}$ frequencies for the ``model low'' CR spectra of \citet{Padovani18}.

The initial spectra for all other elemental CR species was obtained by multiplying the spectrum of protons by their respective abundances relative to H. Besides H, 29 more chemical elements were considered, in addition to the important $^2$D and $^3$He isotopes, for a total of 32 CR species. The species' abundances are based mostly on \textit{Voyager~1} data and are listed in Table~2 of Paper~\citetalias{K18} (see references therein).

\subsubsection{CR spectra at different column densities}
\label{nhsp}

\begin{figure}[ht!]
\vspace{-2cm}
\hspace{-1.0cm}
	\includegraphics[width=13cm]{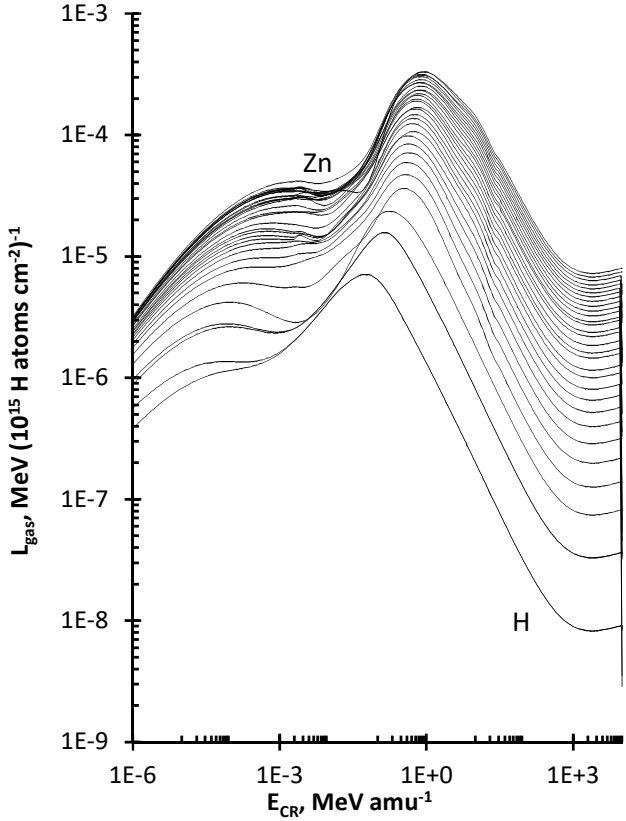}
\caption{Energy loss function $L_{\rm gas}$ for CR species traversing interstellar matter; $E_{\rm lost}$ includes also losses due to ISM atoms other than H.}
\label{fig-losf}
\end{figure}

The CR particles with their initial spectra propagate through the ISM with the column densities $N_{\rm H}$ listed in Table~\ref{tab-nh}. The values of $N_{\rm H}$ indicate only the amount of H atoms traversed by the CR particles, while the medium contains also other elements. Here, the gas was assumed to have the local ISM abundances, consisting of 70.6\,\% H, 28.0\,\% He and 1.4\,\% other elements by mass \citep[][parameter $F_*$ taken to be 1]{Jenkins09}.

\begin{figure}[ht!]
	\includegraphics[width=10cm]{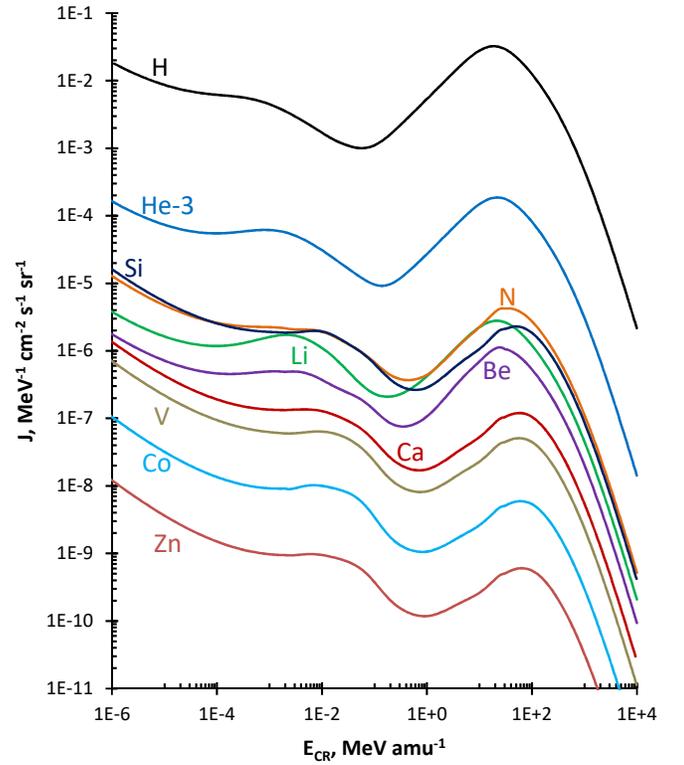}
\caption{Energy spectra for selected cosmic-ray components at $N_{\rm H}=8.80\times10^{22}$\,cm$^{-2}$, corresponding to $A_V=40$\,mag.}
\label{fig-crsp}
\end{figure}
%
\begin{figure}[ht!]
	\includegraphics[width=10cm]{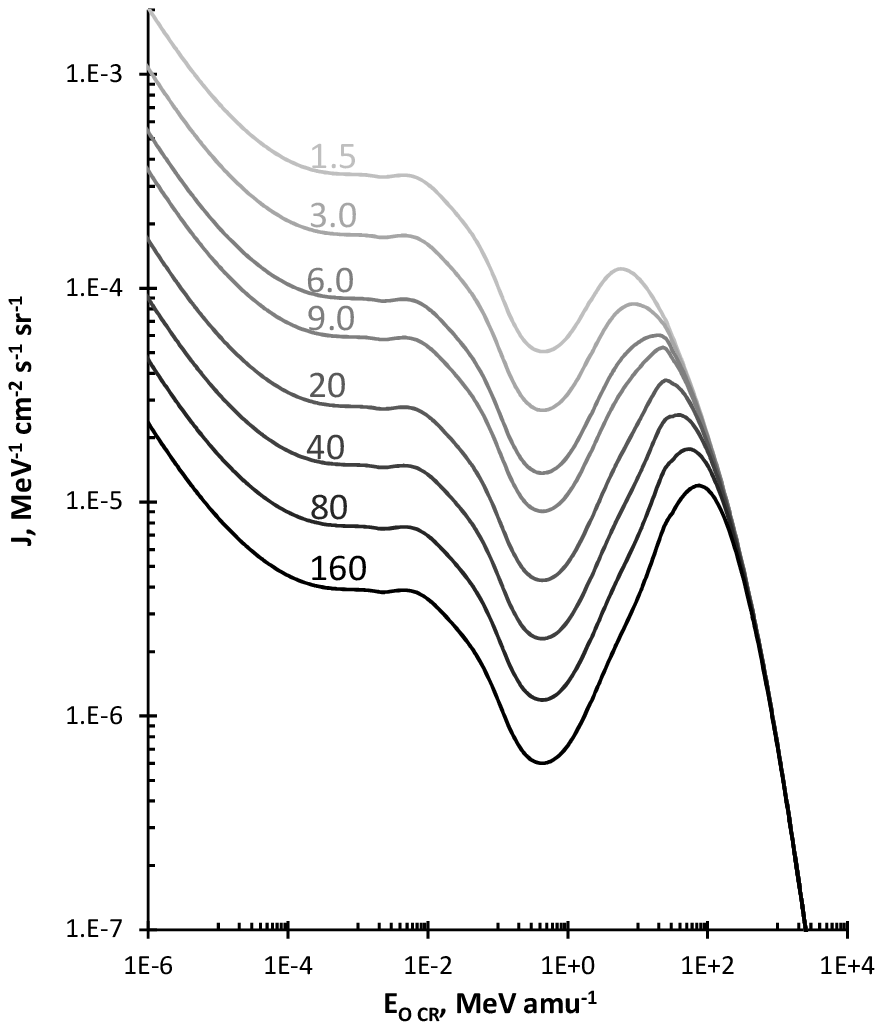}
\caption{Example energy spectra for oxygen CR nuclei at different $A_V$ value for each curve, with $N_{\rm H}/A_V=2.20\times10^{21}$\,cm$^{-2}$.}
\label{fig-Ocr}
\end{figure}

The energy loss functions of the 32 CR species $L_{\rm gas}$ for traversing ISM with the above gas composition were calculated with the \textsc{srim}2013 package \citep{Ziegler10} and are shown in Figure~\ref{fig-losf}. The \textsc{srim} program does not consider CR energy loss via pion production, which becomes important at GeV energies \citep{Padovani18} and the CR spectrum at the maximum $N_{\rm H}/A_V=3.52\times10^{23}$\,cm$^{-2}$ is overestimated by a factor of $\approx2$ (Alexei Ivlev, private communication). This means that the calculated data must not be extrapolated and thus the results are not applicable, for example, to dense circumstellar disks or the central part of massive prestellar cores. However, in objects with such a high $N_{\rm H}$, the efficiency of CRD is negligible. Figure~\ref{fig-crsp} shows an example of the calculated spectra deep into the cloud at an $A_V$ value of 40\,mag for several CR species. Figure~\ref{fig-Ocr} shows an example with oxygen nuclei how the CR spectra changes with increasing $A_V$.

\subsection{CR energy loss in grains}
\label{grloss}

\begin{figure}[ht!]
\vspace{3cm}
\hspace{-1.8cm}
	\includegraphics[width=12cm]{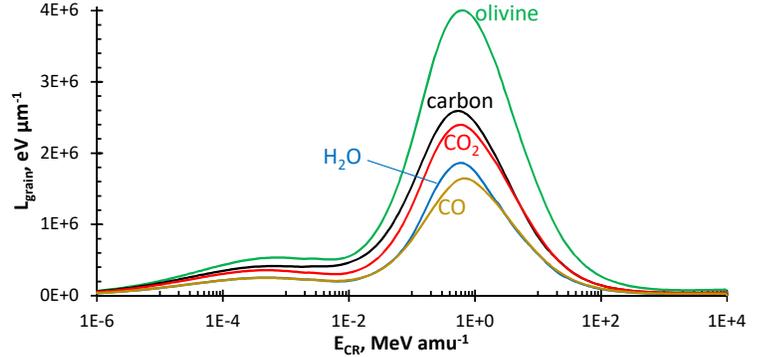}
\vspace{-4cm}
\caption{Energy loss functions $L_{\rm grain}$ for magnesium CR nuclei traversing different grain materials.}
\label{fig-losf2}
\end{figure}

When quantifying the interaction of CRs with interstellar grains, it was assumed that CRs may arrive from all directions with equal probability, i.e., the spectra (per steradian) was multiplied by a factor of $4\pi$, similarly to Paper~\citetalias{K18}. This means that the results are most relevant for grains residing in the center of a spherical cloud. In other geometric setups, the calculated frequencies $f_E$ (for CR impacts giving grains a certain amount of energy $E_{\rm grain}$) and $f_T$ (for impacts raising grain temperature to a certain $T_{\rm CR}$) may have to be modified.

The energy spectra of grains hit by CRs characterizes how often a grain receives a certain amount of energy. These data are less directly applicable but are more general. For example, they may allow recalculating the temperature spectra of CR-heated grains with different material heat capacity or density, or estimating the chemical processing rate of ices.

When a CR particle with an energy $E_{\rm CR}$ hits the grain, the particle loses part of its energy, $E_{\rm lost}$. Generally, $E_{\rm lost}=sL_{\rm grain}$, where $s$ is the track length through the refractory core ($s_{\rm core}$) and ice ($s_{\rm ice}$) parts of the grain and $L_{\rm grain}$ is the aggregate energy loss function in the grain for a given CR nucleus, made up from the loss functions for the separate grain materials. For low-energy particles that come to a complete stop in the grain $E_{\rm CR}=E_{\rm lost}$. In a 0.1\,$\mu$m bare grain, such stoppage occurs for ions with total energies up to 20\,keV for protons to about 230\,keV for iron group nuclei.

The energy loss functions $L_{\rm grain}$ for the 32 CR nuclei were calculated with the \textsc{srim}2013 package separately for each of the five constituent materials -- olivine, carbon, and H$_2$O, CO$_2$, and CO ices. Figure~\ref{fig-losf2} shows an example for the loss functions, in this case for Mg nuclei in grain constituent materials.

We assume that olivine and carbon are homogeneously mixed in the grain core, while the three ice components are homogeneously mixed in the icy mantle. Thus, the energy loss functions for both the grain core and the mantle were obtained as the weighed average of the $L_{\rm grain}$ for the respective chemical components of the core and mantle. The proportions of these components are given in Table~\ref{tab-nh}. The energy $E_{\rm lost}$ for the whole grain was then obtained by adding the CR energies lost in the grain core and the mantle. The CR track lengths $s$ were taken from the grain models described in Section~\ref{mdl}.

\section{Results}
\label{rslt}

\subsection{CR-heated grain energy spectra}
\label{rslt-E}

\begin{figure*}[ht!]
\gridline{\hspace{-0.5cm}
          \fig{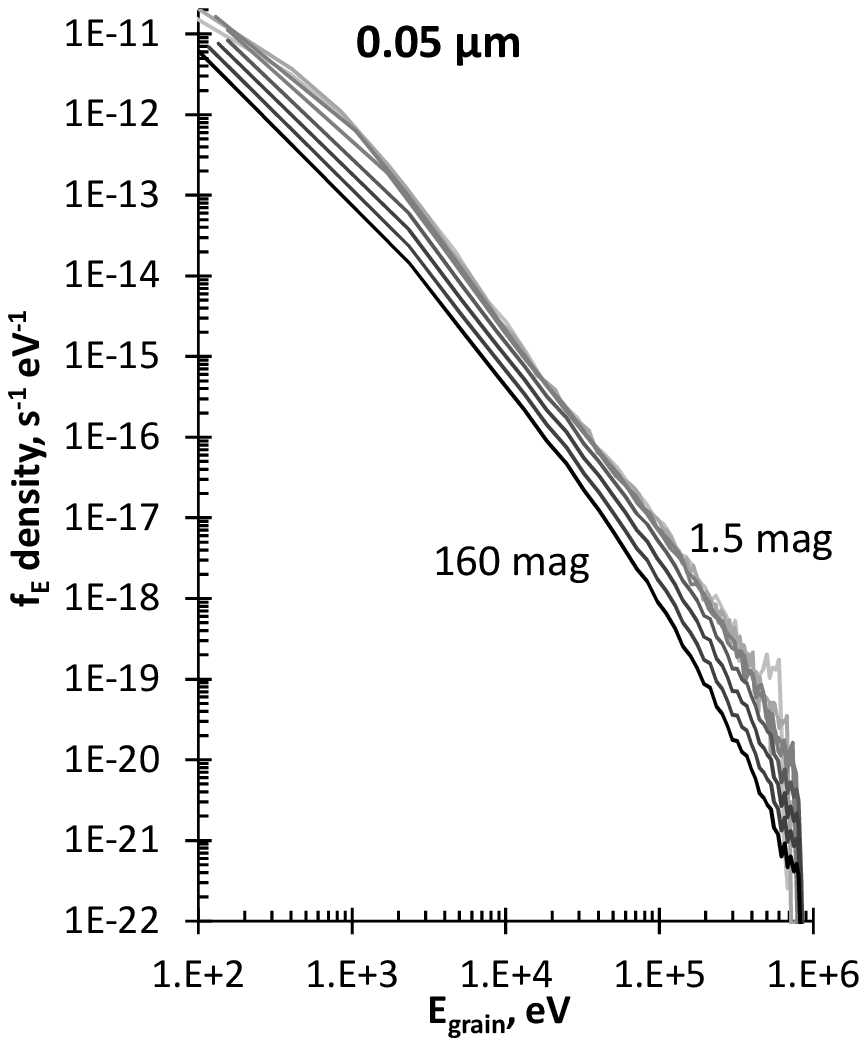}{0.4\textwidth}{}
          \hspace{-1.5cm}
					\fig{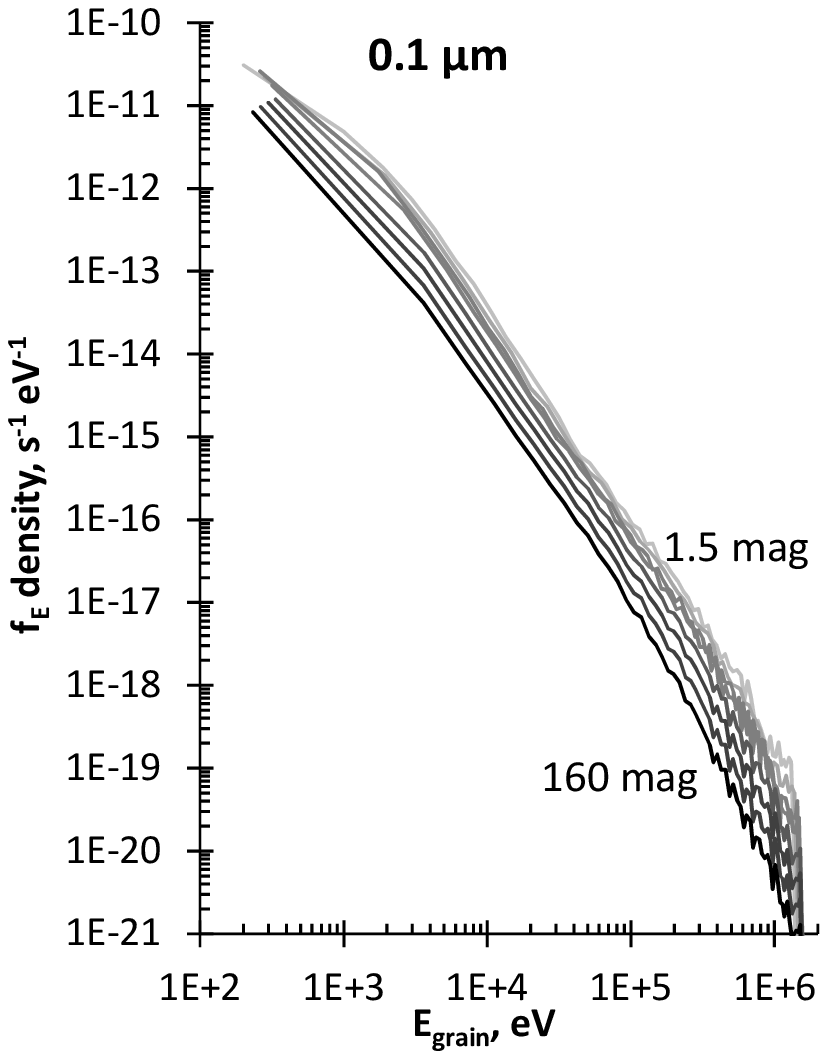}{0.4\textwidth}{}
          \hspace{-1.5cm}
          \fig{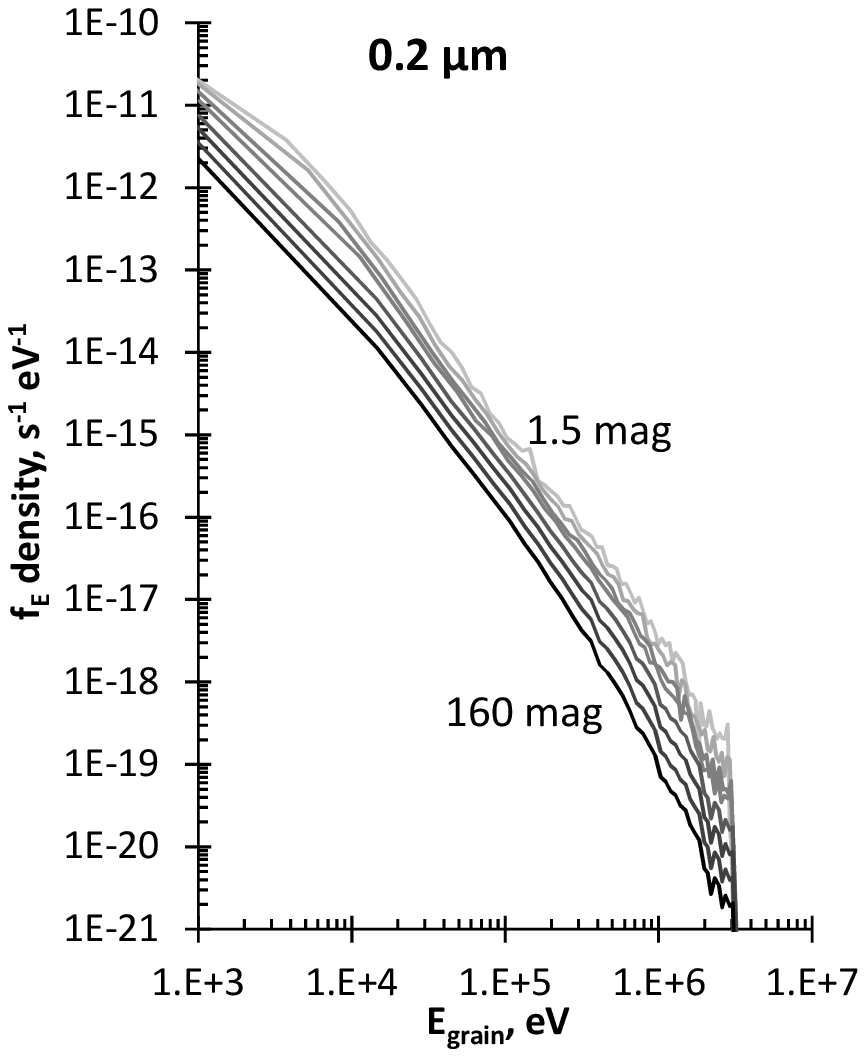}{0.4\textwidth}{}
          }
\caption{Energy $E_{\rm grain}$ spectra for grains with core sizes 0.05, 0.1, and 0.2\,$\mu$m hit by cosmic rays at $A_V$ 1.5; 3.0; 6.0; 9.0; 20; 40; 80; and 160\,mag, with $N_{\rm H}/A_V=2.20\times10^{21}$\,cm$^{-2}$.}
\label{fig-Egr}
\end{figure*}

When the CR particle has deposited the energy $E_{\rm lost}$ in the grain, a part of this energy can leave the grain with fast electrons. The remaining part $x_E$ was estimated with analytical functions derived from the data in Table~B1 of \citet{Leger85}. The resulting energy $E_{\rm grain}=E_{\rm lost}x_E$ was then assumed to thermalize within the grain, raising its temperature to $T_{\rm CR}$. The $E_{\rm grain}$ frequency spectra for all grain types are listed in Appendix~\ref{app-b}.

The procedure described above in Section~\ref{mthd} leaves us with a grain that has just received thermal energy $E_{\rm grain,1}(Z,E_{\rm CR},\rm track)$ (subscript `1' indicates a single CR impact event), which depends on the impacting CR element $Z$ (32 possible values), its energy $E_{\rm CR}$ (261--308 values per element, according to \textsc{srim} tables), and the type of track along which the CR particle transits the grain (3--5 possibilities per grain type). An $E_{\rm grain,1}$ entry includes also its associated frequency $f_{E,1}$. $E_{\rm grain,1}$ is coupled to the respective temperature of a grain $T_{\rm CR,1}$ with the following relation:
\begin{equation}
    E_{\rm grain} = \int^{T_{\rm CR}}_{10}C {\rm d} T\,.
	\label{rslt1}
\end{equation}
All the $T_{\rm CR,1}$ obtained from the $E_{\rm grain,1}$ array via Equation~(\ref{rslt1}) for a grain type were combined in a single list and arranged according to their values. (The entries also include also $E_{\rm grain,1}$ and frequency values, with the latter renamed $f_{T,1}$.) To produce the temperature and its conjugated grain energy spectra, entries with similar $T_{\rm CR,1}$ values across a specific temperature $\Delta T_{\rm CR}$ and energy $\Delta E_{\rm grain}$ interval were grouped together (see also below section). The corresponding $T_{\rm CR}$ was obtained as a weighed average:
\begin{equation}
    T_{\rm CR} = \frac{\sum T_{\rm CR,1} f_{T,1}}{\sum f_{T,1}}\,,
	\label{rslt2}
\end{equation}
while the combined frequency is simply the sum of the frequencies of the separate entries $f_T=\sum f_{T,1}$.  A number of $E_{\rm grain}$ values with their corresponding $f_E$ constitute the energy spectrum of CR-hit grains, while $T_{\rm CR}$ with $f_T$ constitute the temperature spectrum for whole-grain heating. In this work, for a certain grain type we employ only $\Delta T_{\rm CR}$ and $\Delta E_{\rm grain}$ that are mutually interchangeable via relation~(\ref{rslt1}), which means that $f_T$ and $f_E$ always are equal for the given interval.

The energy spectra for a single-size grain is characterized by two trends with increasing $A_V$. First, the overall $E_{\rm grain}$ increases because of the longer tracks in thicker icy mantles. Second, the overall frequency $f_E$ for CR impacts by particles of certain species and energy $E$ decreases because of absorption of CRs in the cloud. Figure~\ref{fig-Egr} shows that these effects partially cancel each other with the result that the frequency density for 0.05\,$\mu$m grains at a given energy is similar for $A_V$ values up to 9.0\,mag, where active ice mantle growth is assumed. For the 0.1 and 0.2\,$\mu$m grains, this canceling of effects is limited because the ice layer contributes relatively much less to the increase of CR track length in the grain.

\subsection{CR-heated grain temperature spectra}
\label{rslt-T}

\begin{figure*}[ht!]
\gridline{\hspace{-0.5cm}
          \fig{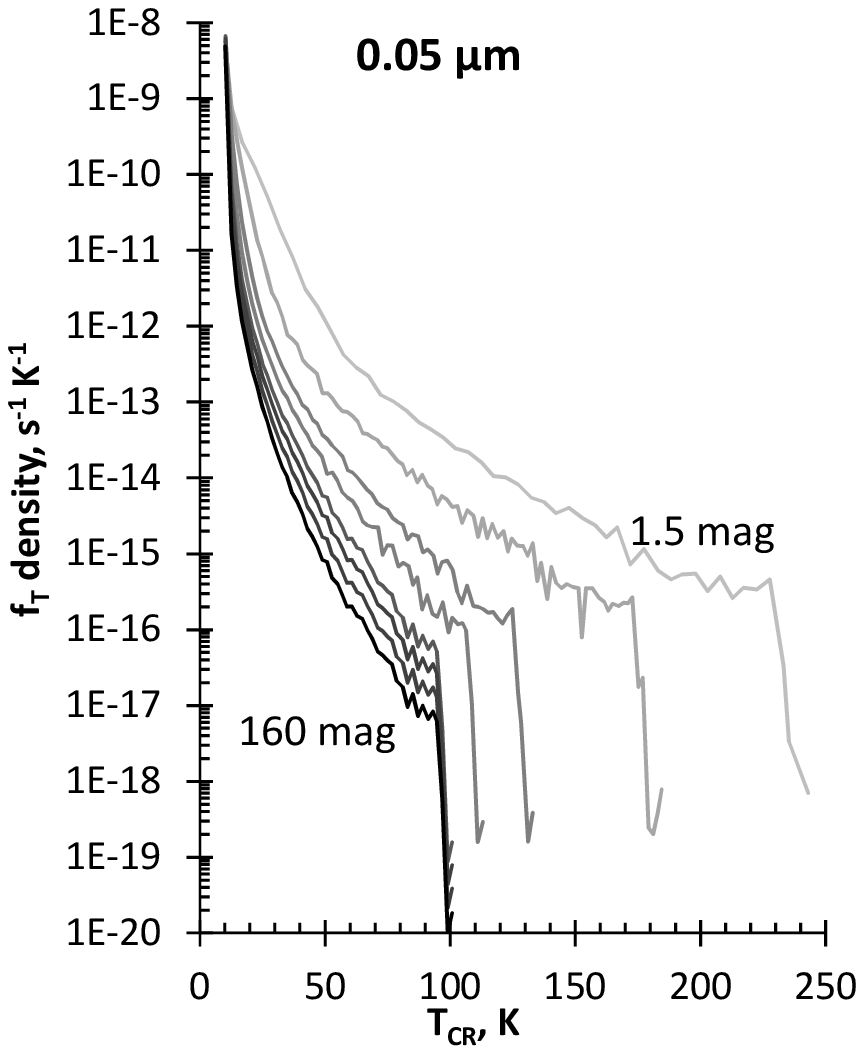}{0.4\textwidth}{}
          \hspace{-1.5cm}
					\fig{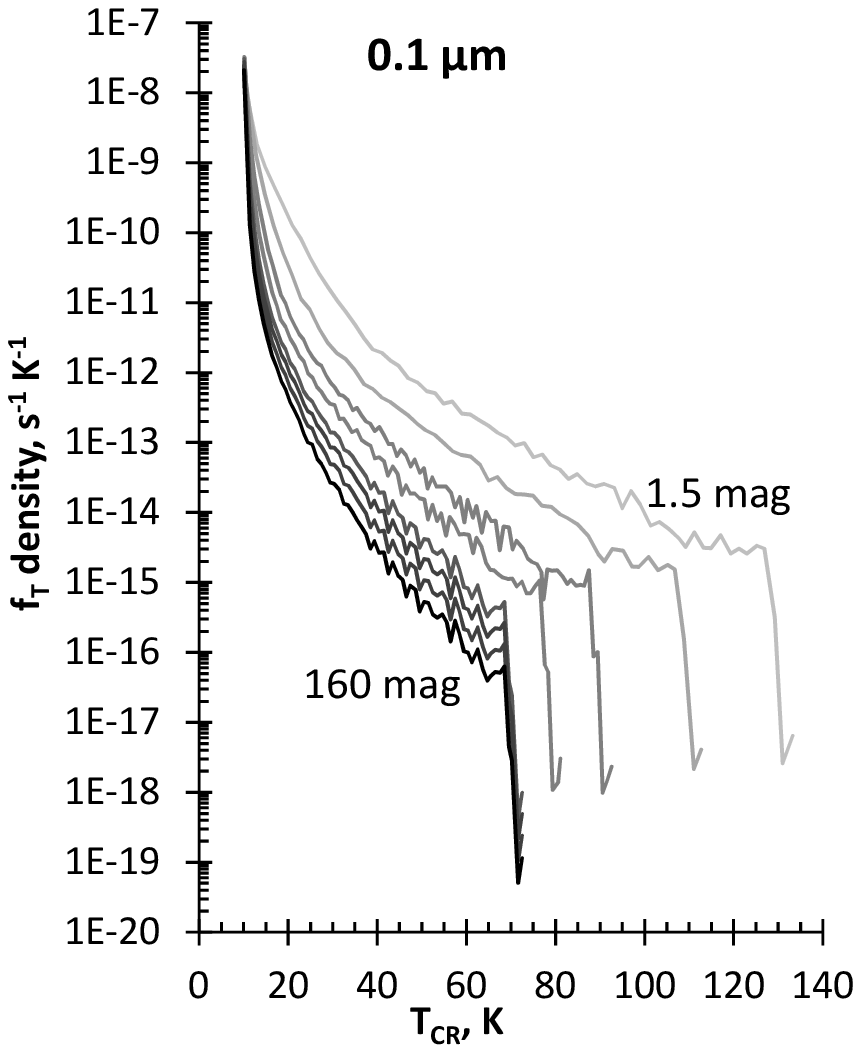}{0.4\textwidth}{}
          \hspace{-1.5cm}
          \fig{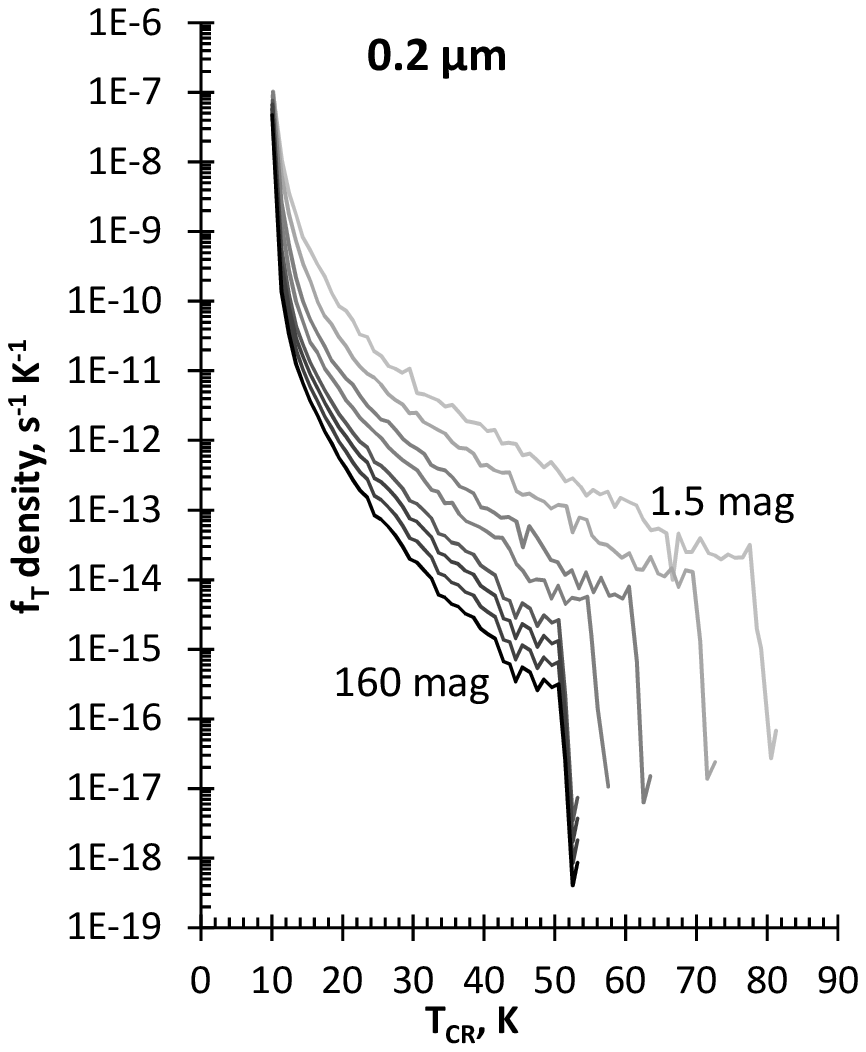}{0.4\textwidth}{}
          }
\caption{Temperature spectra for grains with core sizes 0.05, 0.1, and 0.2\,$\mu$m hit by cosmic rays at $A_V$ 1.5; 3.0; 6.0; 9.0; 20; 40; 80; and 160\,mag, with $N_{\rm H}/A_V=2.20\times10^{21}$\,cm$^{-2}$.}
\label{fig-Tcr}
\end{figure*}
Within about $10^{-9}$\,s after the CR impact, the grain obtains an uniform temperature \citep{Leger85,Bringa04}. The highest $T_{\rm CR}$ of 245\,K is achieved by the smallest -- 0.05\,$\mu$m bare grains (at $A_V=1.5$\,mag), while 0.05\,$\mu$m grains with a 0.01\,$\mu$m ice mantle reach 185\,K. Sublimation of the hypervolatiles from such grains starts at timescales of $10^{-10}$\,s \citep{Kalvans20}. This means that sublimation and some grain cooling may occur, when the heat has not yet fully dissipated within the grain, and the high peak temperatures actually are not reached. This means that the peak temperatures exceeding $\approx150$\,K in our data are not entirely correct. This discrepancy does not affect the usability of our results in estimating CRD yield because at temperatures higher than $\approx40$\,K, the yield depends on the grain's thermal energy content $E_{\rm grain}$, independently of the exact value of $T_{\rm CR}$. Also, observations and astrochemical models indicate that the bare or thin-ice-covered grains at low extinctions are poor with the hypervolatile ices.

Figure~\ref{fig-Tcr} shows graphically the principal output of this study: the temperature spectra of CR-heated grains at different cloud depths, taking into account grain growth due to accumulation of icy mantles. The grains achieve an initial peak temperature $T_{\rm CR}$ with a frequency $f_T$. These data are listed in numerical form in Appendix~\ref{app-b}. For grains with $T_{\rm CR}$ in excess of 200\,K, the $\Delta T_{\rm CR}$ interval for data points was taken to be 5\,K, for $T_{\rm CR}>100$\,K the interval was taken to be 2\,K, while for $T_{\rm CR}<100$\,K, a 1\,K interval was used. The grain energies $E_{\rm grain}$ were listed corresponding to each $T_{\rm CR}$ value. As stated in Section~\ref{grloss}, we consider grains in an spherically symmetric environment with an uniform CR intensity from all directions.

\begin{table*}
\centering
\caption{Frequency $f_T$, s$^{-1}$, of CR impacts that lift the temperature of 0.05\,$\mu$m grains above a minimum $T_{\rm CR}$ threshold.}
\label{tab-sum.05}
\begin{tabular}{lcccccccc}
\tablewidth{0pt}
\hline
\hline
 & \multicolumn{8}{c}{$A_V=N_H/(2.2\times10^{21})$} \\
$T_{\rm CR}$, K & 1.5 & 3.0 & 6.0 & 9.0 & 20 & 40 & 80 & 160 \\
\hline
$>$20 & 1.03E-09 & 1.28E-10 & 2.26E-11 & 1.07E-11 & 5.17E-12 & 3.42E-12 & 2.18E-12 & 1.34E-12 \\
$>$30 & 1.69E-10 & 1.52E-11 & 3.43E-12 & 1.64E-12 & 7.25E-13 & 4.39E-13 & 2.52E-13 & 1.36E-13 \\
$>$40 & 3.61E-11 & 4.51E-12 & 1.02E-12 & 4.45E-13 & 1.64E-13 & 9.14E-14 & 4.84E-14 & 2.44E-14 \\
$>$50 & 1.17E-11 & 1.93E-12 & 3.81E-13 & 1.41E-13 & 4.65E-14 & 2.46E-14 & 1.25E-14 & 6.14E-15 \\
$>$60 & 5.33E-12 & 9.64E-13 & 1.55E-13 & 5.27E-14 & 1.51E-14 & 7.76E-15 & 3.86E-15 & 1.87E-15 \\
$>$70 & 2.79E-12 & 5.02E-13 & 6.78E-14 & 2.06E-14 & 4.47E-15 & 2.27E-15 & 1.12E-15 & 5.43E-16 \\
$>$80 & 1.66E-12 & 2.65E-13 & 3.14E-14 & 7.78E-15 & 1.33E-15 & 6.73E-16 & 3.32E-16 & 1.60E-16 \\
$>$90 & 1.00E-12 & 1.49E-13 & 1.45E-14 & 2.79E-15 & 3.63E-16 & 1.83E-16 & 8.99E-17 & 4.33E-17 \\
$>$100 & 6.19E-13 & 8.83E-14 & 6.14E-15 & 9.77E-16 & 2.08E-19 & 1.04E-19 & 5.11E-20 & 2.44E-20 \\
$>$150 & 7.74E-14 & 5.96E-15 &  &  &  &  &  &  \\
$>$200 & 1.14E-14 &  &  &  &  &  &  &  \\
\hline
\end{tabular}
\end{table*}
%
\begin{table*}
\centering
\caption{Frequency $f_T$, s$^{-1}$, of CR impacts that lift the temperature of 0.1\,$\mu$m grains above a minimum $T_{\rm CR}$ threshold.}
\label{tab-sum0.1}
\begin{tabular}{lcccccccc}
\tablewidth{0pt}
\hline
\hline
 & \multicolumn{8}{c}{$A_V=N_H/(2.2\times10^{21})$} \\
$T_{\rm CR}$, K & 1.5 & 3.0 & 6.0 & 9.0 & 20 & 40 & 80 & 160 \\
\hline
$>$20 & 6.62E-10 & 1.19E-10 & 2.97E-11 & 1.43E-11 & 6.22E-12 & 4.01E-12 & 2.48E-12 & 1.48E-12 \\
$>$30 & 7.56E-11 & 1.64E-11 & 4.46E-12 & 2.02E-12 & 7.65E-13 & 4.42E-13 & 2.40E-13 & 1.24E-13 \\
$>$40 & 1.95E-11 & 4.77E-12 & 1.15E-12 & 4.34E-13 & 1.30E-13 & 6.89E-14 & 3.49E-14 & 1.71E-14 \\
$>$50 & 7.20E-12 & 1.68E-12 & 3.30E-13 & 1.12E-13 & 2.93E-14 & 1.50E-14 & 7.42E-15 & 3.59E-15 \\
$>$60 & 3.11E-12 & 6.44E-13 & 1.11E-13 & 2.97E-14 & 4.93E-15 & 2.49E-15 & 1.23E-15 & 5.92E-16 \\
$>$70 & 1.34E-12 & 2.57E-13 & 2.98E-14 & 6.47E-15 & 2.55E-17 & 1.28E-17 & 6.25E-18 & 2.99E-18 \\
$>$80 & 6.12E-13 & 9.99E-14 & 9.07E-15 & 2.26E-18 &  &  &  &  \\
$>$90 & 3.01E-13 & 3.68E-14 & 4.72E-18 &  &  &  &  &  \\
$>$100 & 1.16E-13 & 1.45E-14 &  &  &  &  &  &  \\
\hline
\end{tabular}
\end{table*}
%
\begin{table*}
\centering
\caption{Frequency $f_T$, s$^{-1}$, of CR impacts that lift the temperature of 0.2\,$\mu$m grains above a minimum $T_{\rm CR}$ threshold.}
\label{tab-sum0.2}
\begin{tabular}{lcccccccc}
\tablewidth{0pt}
\hline
\hline
 & \multicolumn{8}{c}{$A_V=N_H/(2.2\times10^{21})$} \\
$T_{\rm CR}$, K & 1.5 & 3.0 & 6.0 & 9.0 & 20 & 40 & 80 & 160 \\
\hline
$>$20 & 3.07E-10 & 1.02E-10 & 3.52E-11 & 1.69E-11 & 7.17E-12 & 4.38E-12 & 2.53E-12 & 1.38E-12 \\
$>$30 & 4.21E-11 & 1.52E-11 & 4.27E-12 & 1.77E-12 & 5.71E-13 & 3.05E-13 & 1.55E-13 & 7.66E-14 \\
$>$40 & 1.10E-11 & 3.13E-12 & 6.96E-13 & 2.50E-13 & 5.68E-14 & 2.89E-14 & 1.43E-14 & 6.91E-15 \\
$>$50 & 2.86E-12 & 7.28E-13 & 1.06E-13 & 3.05E-14 & 2.85E-15 & 1.43E-15 & 7.03E-16 & 3.38E-16 \\
$>$60 & 7.75E-13 & 1.46E-13 & 8.57E-15 &  &  &  &  &  \\
$>$70 & 2.02E-13 & 1.34E-15 &  &  &  &  &  &  \\
$>$80 & 5.94E-17 &  &  &  &  &  &  &  \\
\hline
\end{tabular}
\end{table*}

Tables~\ref{tab-sum.05}--\ref{tab-sum0.2} show summarized output -- the frequency of CR-induced whole-grain heating events that result in the grains exceeding temperature thresholds starting with 20\,K and higher. These data can be directly applied in astrochemical modeling.

\subsection{Summary}
\label{smmr}

In Appendix~\ref{app-b} the temperature and energy spectra for interstellar grains affected by CR hits is given in numerical form. When applying these data in describing sublimation from 0.05\,$\mu$m grains having up to 0.02\,$\mu$m thick icy mantle one has to consider that these grains may exceed the 134\,K critical temperature of CO ice (Section~\ref{chm}). If CO ice is abundant in the mantle, $T_{\rm CR}>134$\,K can be expected to affect the nature of sublimation. Another issue for the 0.05\,$\mu$m grains having a thin icy mantle of $\leq0.1\,\mu$m exceeding temperatures of $\approx150$\,K because the sublimation timescale for CO is comparable to or shorter than the $\approx10^{-9}$\,s$^{-1}$ necessary for heat dissipation throughout the grains. This effect will have little effect because the number of sublimated molecules primarily depends on grain thermal energy, not its exact temperature (Section~\ref{rslt-T}).

\begin{figure*}[ht!]
\plotone{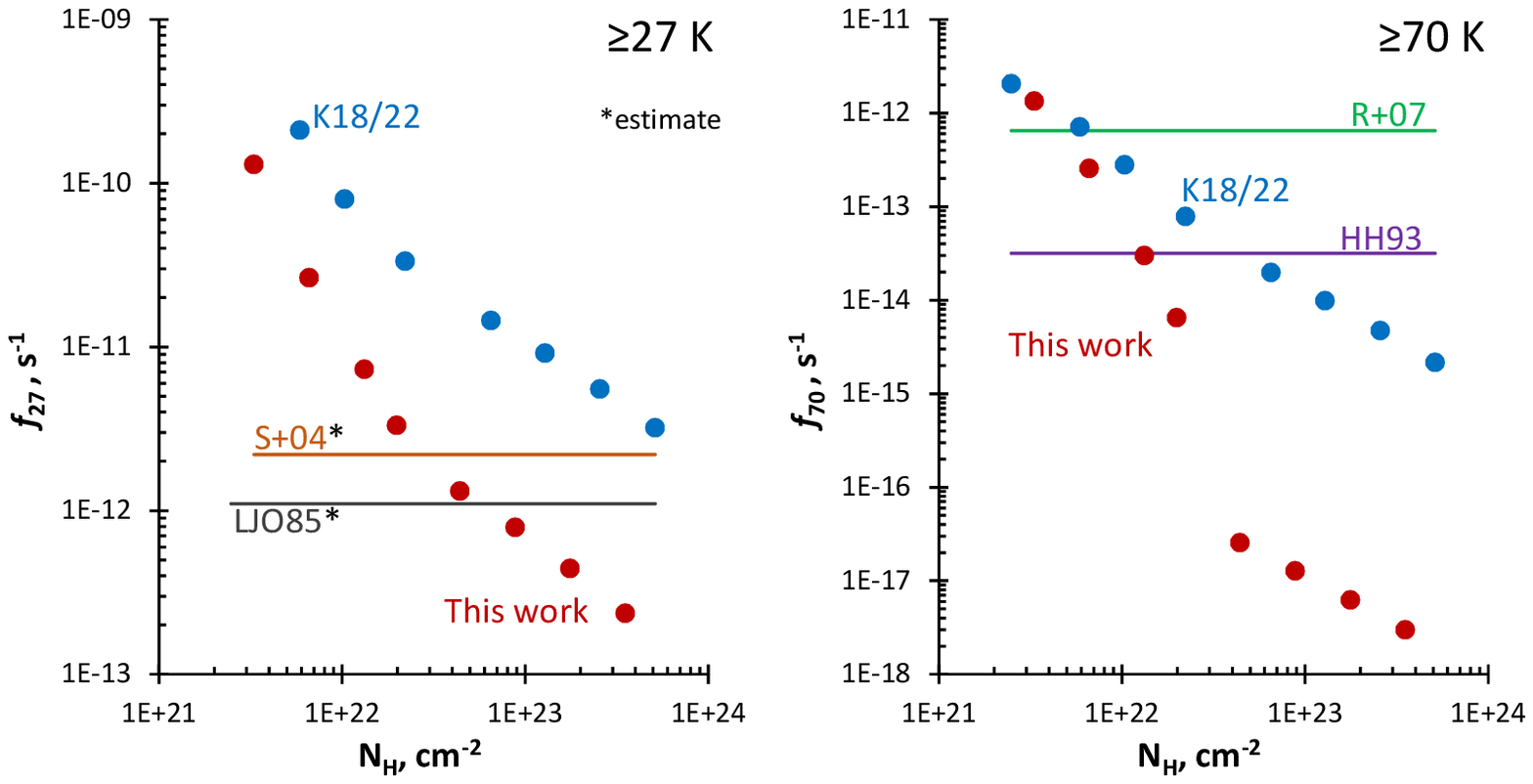}
\caption{Comparison of CR-induced whole-grain heating frequency $f_T$ for $T_{\rm CR}$ above 27\,K and 70\,K by different authors. References: K18/22: Paper~\citetalias{K22}; HH93: \citet{Hasegawa93}; R+07: \citet{Roberts07}; LJO85: \citet{Leger85}; S+04: \citet{Shen04}.}
\label{fig-fT}
\end{figure*}

The data in this paper covers a similar range of $A_V$ and $N_{\rm H}$ values to those in Papers \citetalias{K18} and \citetalias{K22}. Differences include that here we consider a similar ice thickness for all grain sizes (Section~\ref{mdl}). Moreover, here we avoid calculation of $C$ for the grain core with the Debye method, which underestimates $C$ at temperatures below 70\,K and overestimates it at temperatures above \citep{Kalvans21}. The consideration of different components in mixed-composition grains induced two additional changes, allowing for a more accurate data. First, considering the less-dense aC mixed with a silicate material reduced $L_{\rm grain}$ (i.e., the grain receives less energy), compared to that calculated in Paper~\citetalias{K22}. Second, the explicit consideration of carbon oxides produces a higher ice $C$, important for grains with thick mantles abundant with CO and CO$_2$. As a result, for most heating events of 0.2\,$\mu$m grains, which rarely exceed 70\,K, $C$ for grain core and ice is much higher than that calculated in Paper~\citetalias{K22}. For the 0.05\,$\mu$m grains, the lower of $C$ for grain cores after 70\,K comes into play and very high $T_{\rm CR}$ can be achieved. However, at high $A_V$, more important is that such small grains with a thick ice layer contain a substantial proportion of carbon oxides, which reduces their ability to reach $T_{\rm CR}$ in excess of 100\,K. For 0.1\,$\mu$m grains, the effects are mixed, resulting to higher $T_{\rm CR}$ for bare grains and lower $T_{\rm CR}$ for grains with thick icy mantles, compared to the data of Paper~\citetalias{K22}.

The calculated $E_{\rm grain}$ values are lower than those of Paper~\citetalias{K22} by a factor of $\approx0.8$ The differences between the two calculations are more pronounced for $T_{\rm CR}$ and $f_T$, mainly because of the different $C$. The maximum $T_{\rm CR}$ for 0.1\,$\mu$m grains with a 0.03\,$\mu$m ice layer is 72.5\,K in this study and 91.3\,K in Paper~\citetalias{K22} data. The corresponding frequencies $f_{70}$ are not directly comparable because of the different $N_H$ values in the two studies. However, it can be said that for grains with a thick $\approx0.03$\,$\mu$m ice layer, this study predicts an order of magnitude lower $T_{\rm CR}$ than that of Paper~\citetalias{K22} for temperatures in excess of 30\,K. Lower $T_{\rm CR}$ generally means that large 0.1 or 0.2\,$\mu$m grains with thick icy mantles deep into a dense cloud core can relatively rarely exceed the critical 40\,K $T_{\rm CR}$ threshold, where evaporative cooling starts dominating over radiative cooling, and CRD can be efficient. For bare and thinly-iced grains, the frequencies $f_T$ are comparable with Paper~\citetalias{K22}, while $T_{\rm CR}$ here are significantly higher (for example, 133.3\,K versus 103.3\,K in Paper~\citetalias{K22} for 0.1\,$\mu$m bare grains). Figure~\ref{fig-fT} compares the heating frequencies to 27\,K and 70\,K at different $A_V$ to results from previous studies.

\begin{acknowledgments}
JK was been funded by ERDF postdoctoral grant No. 1.1.1.2/VIAA/I/16/194 ``Chemical effects of cosmic-ray induced heating of interstellar dust grains'' being implemented in Ventspils University of Applied Sciences. JRK was funded by ERDF project No. 1.1.1.1/16/A/213 ``Physical and chemical processes in the interstellar medium''. We also thank the Ventspils City Council for support and Kristaps Veitners for some of the figures in the paper. We are also grateful to the anonymous referee, whose comments helped to substantially improve the study. This research has made use of NASA’s Astrophysics Data System.
\end{acknowledgments}

\appendix

\section{Geometrical grain model}
\label{app-a}

\begin{figure*}[ht!]
\gridline{ \hspace{-2cm}
\fig{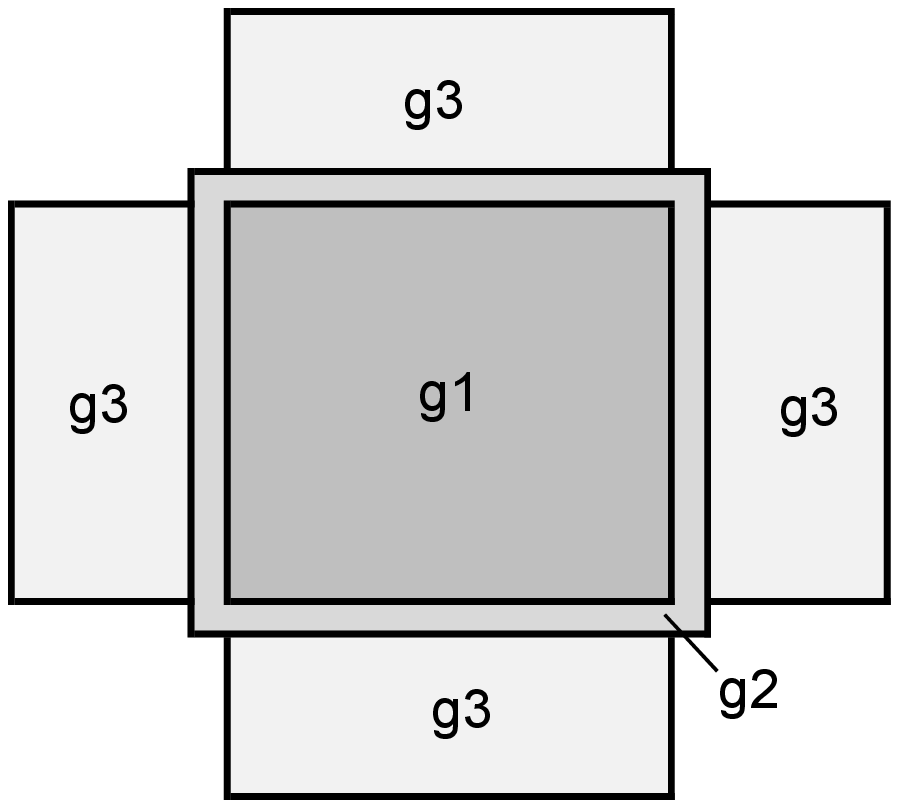}{0.6\textwidth}{(a)} \hspace{-6cm}
\fig{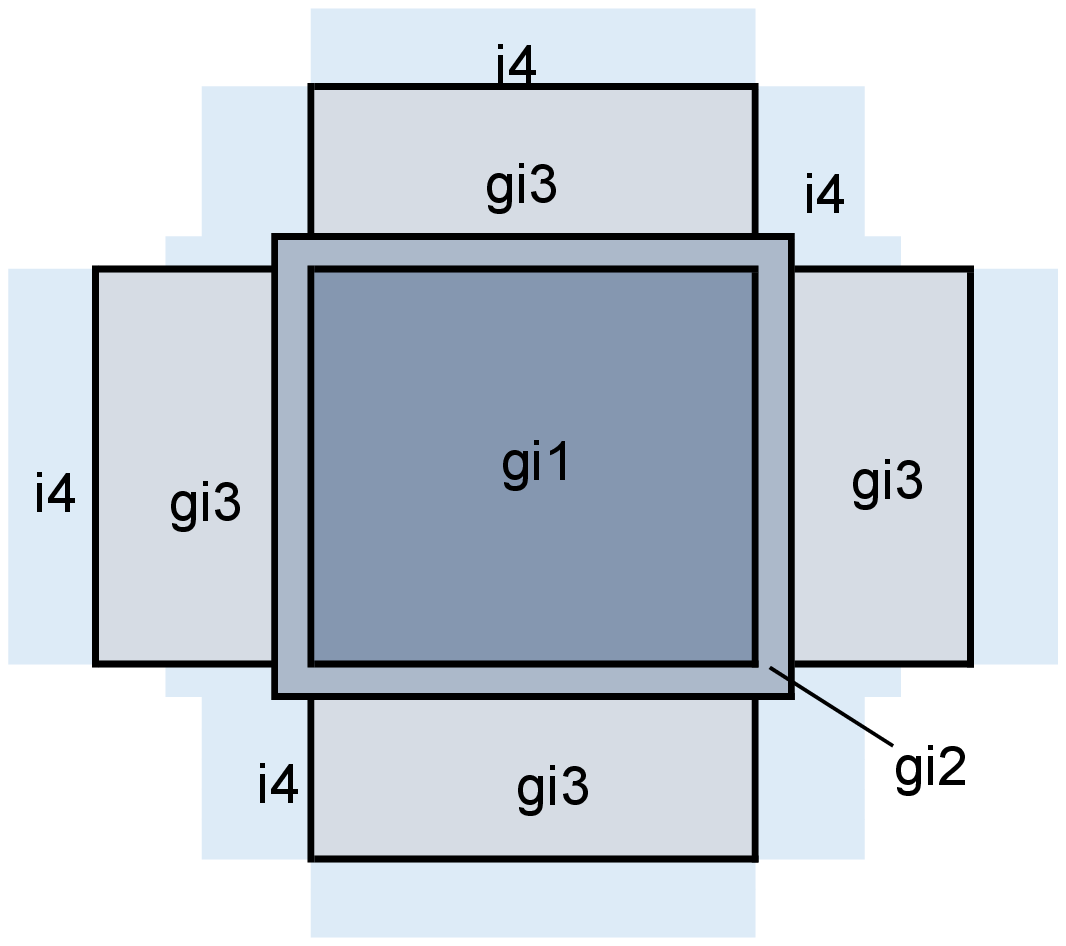}{0.6\textwidth}{(b)} \hspace{-5cm}
\fig{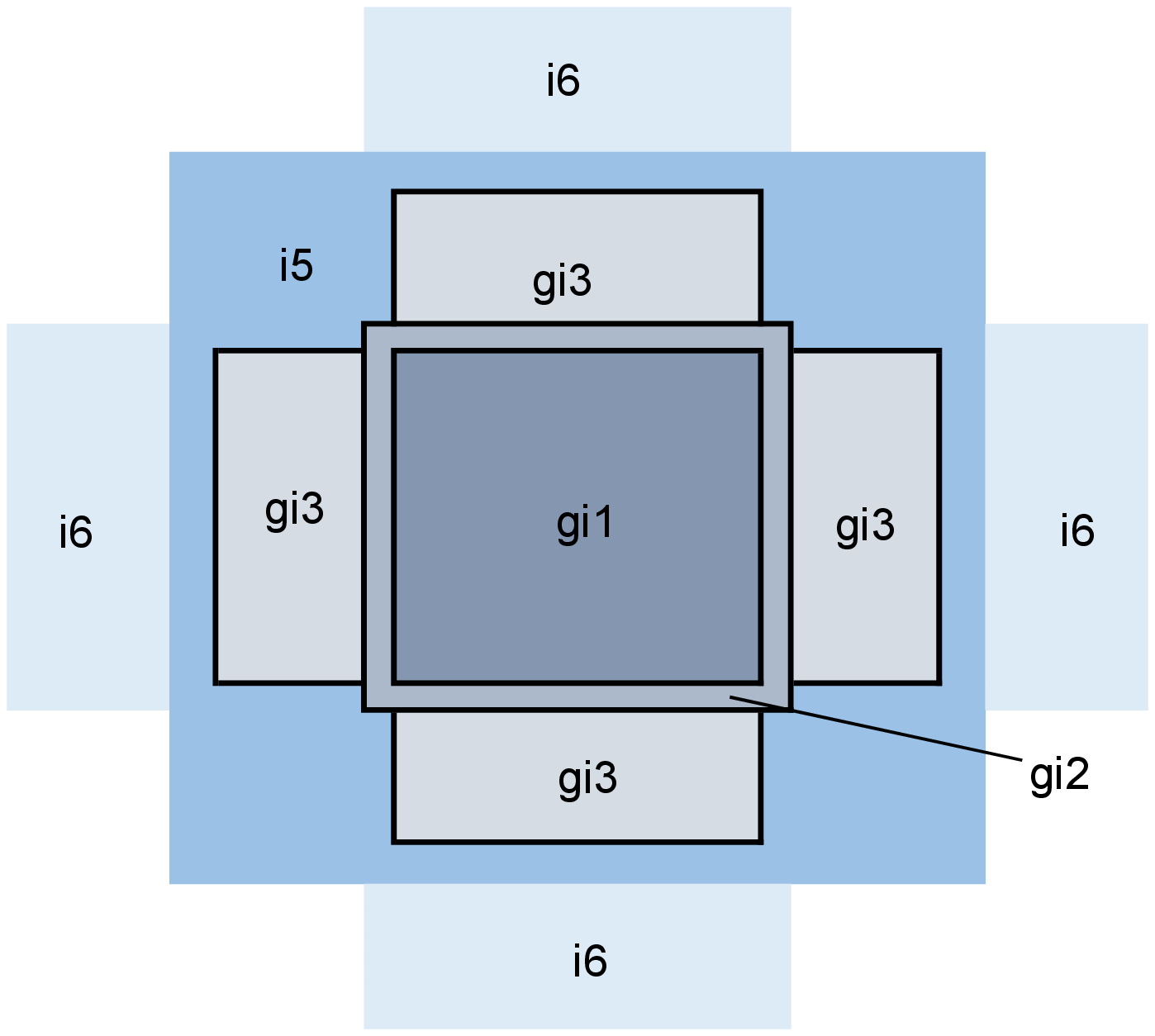}{0.6\textwidth}{(c)} }
\caption{Left to right: cross sections of bare (a) and icy interstellar grain models with thin (b) and thick (c) icy mantles (cf. Figure~\ref{fig-mdl} for bare grain). Different possible CR tracks are identified using the convention `g': track passes only through refractory nucleus of the grain; `i': only through ice; `gi': through ice and grain nucleus.}
\label{fig-cs}
\end{figure*}

Here we offer more details on the parameters for the cuboid grain models employed in the calculations. Figure~\ref{fig-cs} shows the cross-sections of the three geometrical grain types: bare grains (a) (compare with Figure~\ref{fig-mdl}), icy grains (b) and small icy grains (c). For  types (b) and (c), ice thickness can be varied so that the energy $E_{\rm grain}$ and frequency $f_E = f_T$ match those of a spherical grain with a given ice thickness $B$. In turn, $E_{\rm grain}$, which determines $T_{\rm CR}$ depends on track length $s$ of a CR particle passing through the grain, while, and frequencies $f_E$ and $f_T$ (equal in Appendix~\ref{app-b}), depend on grain cross-section area $S$. Models (a) and (c) permit CR entrance and have similar cross-sections along all three axis, while for model (b), CR entrance is permitted along one axis only (with its cross-section featured here), in order to attain $S$ and $s$ most similar to those of a spherical grain. Table~\ref{tab-tracks} lists the values for $S$ and $s$ for each track.
%
\begin{table*}
\centering
\caption{Cross-section $S$, and track lengths through grain core ($s_{\rm core}$) and its icy mantle ($s_{\rm ice}$) of tracks for all grain types (one per line) considered. See Table~\ref{tab-nh} for the relevant $A_V$ and $N_H$ for each ice thickness $B$ and consult Figure~\ref{fig-cs} for the placement of the CR tracks in the grain models.}
\label{tab-tracks}
\begingroup
\footnotesize
\begin{tabular}{ll|ccc|ccc|ccc|cc|cc}
\tablewidth{0pt}
\hline
\hline
$a$, & $B$, & $S$, & $s_{\rm core}$ & $s_{\rm ice}$ & $S$, & $s_{\rm core}$ & $s_{\rm ice}$ & $S$, & $s_{\rm core}$ & $s_{\rm ice}$ & $S$, & $s_{\rm ice}$ & $S$, & $s_{\rm ice}$ \\
$\mu$m & $\mu$m & $\mu{\rm m}^2$ & $\mu$m & $\mu$m & $\mu{\rm m}^2$ & $\mu$m & $\mu$m & $\mu{\rm m}^2$ & $\mu$m & $\mu$m & $\mu{\rm m}^2$ & $\mu$m & $\mu{\rm m}^2$ & $\mu$m \\
\hline
\multicolumn{15}{c}{Grain model (a)\tablenotemark{a}} \\
\multicolumn{2}{c}{CR track} & \multicolumn{3}{c}{g1} & \multicolumn{3}{c}{g2} & \multicolumn{3}{c}{g3} &  &  &  &  \\
0.05 &  & 0.00260 & 0.100 & 0 & 0.000880 & 0.0590 & 0 & 0.00418 & 0.0510 & 0 &  &  &  &  \\
0.1 &  & 0.0104 & 0.200 & 0 & 0.00352 & 0.118 & 0 & 0.0167 & 0.102 & 0 &  &  &  &  \\
0.2 &  & 0.0416 & 0.400 & 0 & 0.0141 & 0.236 & 0 & 0.0669 & 0.204 & 0 &  &  &  &  \\
\hline
\multicolumn{15}{c}{Grain model (b)} \\
\multicolumn{2}{c}{CR track} & \multicolumn{3}{c}{gi1} & \multicolumn{3}{c}{gi2} & \multicolumn{3}{c}{gi3} & \multicolumn{2}{c}{i4} &  \\
0.05 & 0.01 & 0.00260 & 0.100 & 0.0200 & 0.000880 & 0.0590 & 0.0298 & 0.00418 & 0.0510 & 0.0298 & 0.00350 & 0.0510 &  &  \\
0.1 & 0.01 & 0.0104 & 0.200 & 0.0200 & 0.00352 & 0.118 & 0.0250 & 0.0167 & 0.102 & 0.0250 & 0.00659 & 0.102 &  &  \\
0.1 & 0.02 & 0.0104 & 0.200 & 0.0400 & 0.00352 & 0.118 & 0.0596 & 0.0167 & 0.102 & 0.0596 & 0.0140 & 0.102 &  &  \\
0.1 & 0.025 & 0.0104 & 0.200 & 0.0500 & 0.00352 & 0.118 & 0.0822 & 0.0167 & 0.102 & 0.0822 & 0.0177 & 0.102 &  &  \\
0.1 & 0.03 & 0.0104 & 0.200 & 0.0600 & 0.00352 & 0.118 & 0.108 & 0.0167 & 0.102 & 0.108 & 0.0216 & 0.102 &  &  \\
0.2 & 0.01 & 0.0416 & 0.400 & 0.0200 & 0.0141 & 0.236 & 0.0225 & 0.0669 & 0.204 & 0.0225 & 0.0129 & 0.204 &  &  \\
0.2 & 0.02 & 0.0416 & 0.400 & 0.0400 & 0.0141 & 0.236 & 0.0500 & 0.0669 & 0.204 & 0.0500 & 0.0264 & 0.204 &  &  \\
0.2 & 0.025 & 0.0416 & 0.400 & 0.0500 & 0.0141 & 0.236 & 0.0658 & 0.0669 & 0.204 & 0.0658 & 0.0333 & 0.204 &  &  \\
0.2 & 0.03 & 0.0416 & 0.400 & 0.0600 & 0.0141 & 0.236 & 0.0824 & 0.0669 & 0.204 & 0.0824 & 0.0406 & 0.204 &  &  \\
\hline
\multicolumn{15}{c}{Grain model (c)} \\
\multicolumn{2}{c}{CR track} & \multicolumn{3}{c}{gi1} & \multicolumn{3}{c}{gi2} & \multicolumn{3}{c}{gi3} & \multicolumn{2}{c}{i5} & \multicolumn{2}{c}{i6} \\
0.05 & 0.02 & 0.00260 & 0.100 & 0.0400 & 0.000880 & 0.0590 & 0.0430 & 0.004182 & 0.0510 & 0.0510 & 0.00274 & 0.102 & 0.00472 & 0.0590 \\
0.05 & 0.025 & 0.00260 & 0.100 & 0.0500 & 0.000880 & 0.0590 & 0.0516 & 0.004182 & 0.0510 & 0.0596 & 0.00457 & 0.111 & 0.00531 & 0.0590 \\
0.05 & 0.03 & 0.00260 & 0.100 & 0.0600 & 0.000880 & 0.0590 & 0.0605 & 0.004182 & 0.0510 & 0.0685 & 0.00662 & 0.120 & 0.00566 & 0.0590 \\
\hline
\end{tabular}
\endgroup
\tablenotetext{a}{ From Figure~\ref{fig-cs}}
\end{table*}
%

\section{Temperature and energy spectra of interstellar grains hit by cosmic rays.}
\label{app-b}

Here we present the main results of this study -- the temperature and energy spectra of CR-heated mixed olivine-aC grains with radius $a$ covered with mixed H$_2$O-CO$_2$-CO icy mantles (Section~\ref{chm}) in interstellar dark, cold cloud core conditions with initial, ambient grain temperature of 10\,K and gas column density $N_H$ (Table~\ref{tab-nh}). The whole-grain heating temperature $T_{\rm CR}$ (Section~\ref{rslt-T}) is the weighed average for its given $T_{\rm CR}$ interval $\Delta T_{\rm CR}$, which usually spans 1 to 5 K (Equation~(\ref{rslt2})). In a similar manner, the grain's respective thermal energy $E_{\rm grain}$ (Equation~(\ref{rslt2})) is the weighed average of energy interval $\Delta E_{\rm grain}$. CR hits imparting the grains with the energy $E_{\rm grain}$ (and thus heating them to $T_{\rm CR}$) occur with a frequency $f_T$. The latter depends on the adopted CR spectrum (Section~\ref{crsp}). Our primary results, likely more relevant for common interstellar conditions, were calculated with the ``model high'' spectrum from \citet{Padovani18}. As supplemental results, we also provide $f_T$ with the ``model low'' CR spectrum. The initial CR spectra were modified by an $N_H$ that is equal in all directions (spherical symmetry).

%
\startlongtable
\begin{deluxetable}{lcccccccc}
\tabletypesize{\footnotesize}
\tablecaption{
Bare grains with $a=0.05$~$\mu$m, shielded by interstellar gas with $N_H=3.30\times10^{21}$ H\,atoms\,cm$^{-2}$. For temperatures of 134.45\,K and above, the critical point of CO ice is surpassed (see Section~\ref{chm}).
}
\label{tab1_0.05_1.5}
\tablehead{
\colhead{No.} & \colhead{$T_{\rm CR}$, K} & \colhead{interval, K} & \colhead{$\Delta T_{\rm CR}$, K} & \colhead{$E_{\rm grain}$, eV} & \colhead{interval, eV} & \colhead{$\Delta E_{\rm grain}$, eV} & \colhead{$f_T$, s$^{-1}$, High} & \colhead{$f_T$, s$^{-1}$, Low}
}
\startdata
1 & 11.82 & 10.01--15.00 & 5.00 & 7.94E+01 & 0.7905--2.406E+02 & 2.398E+02 & 4.71E-09 & 2.64E-09 \\
2 & 17.12 & 15.01--20.00 & 5.00 & 3.95E+02 & ...--6.455E+02 & 4.049E+02 & 1.30E-09 & 6.86E-10 \\
3 & 22.08 & 20.01--25.00 & 5.00 & 9.05E+02 & ...--1.329E+03 & 6.831E+02 & 6.08E-10 & 3.39E-10 \\
4 & 27.08 & 25.01--30.00 & 5.00 & 1.74E+03 & ...--2.388E+03 & 1.060E+03 & 2.48E-10 & 1.60E-10 \\
5 & 32.07 & 30.01--35.00 & 5.00 & 2.98E+03 & ...--3.913E+03 & 1.524E+03 & 9.26E-11 & 6.17E-11 \\
6 & 37.09 & 35.01--40.00 & 5.00 & 4.73E+03 & ...--5.977E+03 & 2.064E+03 & 4.04E-11 & 2.78E-11 \\
7 & 42.07 & 40.01--45.00 & 5.00 & 7.04E+03 & ...--8.653E+03 & 2.676E+03 & 1.54E-11 & 1.15E-11 \\
8 & 47.13 & 45.01--50.00 & 5.00 & 1.00E+04 & ...--1.199E+04 & 3.333E+03 & 8.97E-12 & 6.78E-12 \\
\enddata
\end{deluxetable}

%
\startlongtable
\begin{deluxetable}{lcccccccc}
\tabletypesize{\footnotesize}
\tablecaption{
Grains with $a=	0.05$~$\mu$m and 0.01\,$\mu$m thick icy mantle, shielded by interstellar gas with $N_H=	6.60\times10^{21}$ H\,atoms\,cm$^{-2}$. For temperatures of 134.45\,K and above, the critical point of CO ice is surpassed (see Section~\ref{chm}).
}
\label{tab2_0.05_3.0}
\tablehead{
\colhead{No.} & \colhead{$T_{\rm CR}$, K} & \colhead{interval, K} & \colhead{$\Delta T_{\rm CR}$, K} & \colhead{$E_{\rm grain}$, eV} & \colhead{interval, eV} & \colhead{$\Delta E_{\rm grain}$, eV} & \colhead{$f_T$, s$^{-1}$, High} & \colhead{$f_T$, s$^{-1}$, Low}
}
\startdata
1 & 10.65 & 10.01--12.00 & 2.00 & 7.68E+01 & 2.073--2.589E+02 & 2.568E+02 & 7.27E-09 & 1.89E-09 \\
2 & 12.84 & 12.01--14.00 & 2.00 & 4.06E+02 & ...--6.346E+02 & 3.757E+02 & 1.39E-09 & 6.16E-11 \\
3 & 14.86 & 14.01--16.00 & 2.00 & 8.46E+02 & ...--1.154E+03 & 5.190E+02 & 5.71E-10 & 1.59E-11 \\
4 & 16.89 & 16.01--18.00 & 2.00 & 1.44E+03 & ...--1.832E+03 & 6.788E+02 & 2.35E-10 & 5.89E-12 \\
5 & 18.88 & 18.01--20.00 & 2.00 & 2.20E+03 & ...--2.694E+03 & 8.614E+02 & 1.13E-10 & 2.92E-12 \\
6 & 20.92 & 20.01--22.00 & 2.00 & 3.17E+03 & ...--3.747E+03 & 1.053E+03 & 5.48E-11 & 1.44E-12 \\
7 & 22.90 & 22.01--24.00 & 2.00 & 4.30E+03 & ...--5.001E+03 & 1.254E+03 & 2.66E-11 & 8.06E-13 \\
8 & 24.92 & 24.01--26.00 & 2.00 & 5.66E+03 & ...--6.461E+03 & 1.460E+03 & 1.67E-11 & 5.25E-13 \\
\enddata
\end{deluxetable}

%
\startlongtable
\begin{deluxetable}{lcccccccc}
\tabletypesize{\footnotesize}
\tablecaption{
Grains with $a=	0.05	$~$\mu$m and 0.020	$\mu$m thick icy mantle,  shielded by interstellar gas with $N_H=	1.32\times10^{22}$ H\,atoms\,cm$^{-2}$. The highest $T_{\rm CR}$ values are close to the critical point of CO ice of 134.45\,K (Section~\ref{chm}).
}
\label{tab3_0.05_6.0}
\tablehead{
\colhead{No.} & \colhead{$T_{\rm CR}$, K} & \colhead{interval, K} & \colhead{$\Delta T_{\rm CR}$, K} & \colhead{$E_{\rm grain}$, eV} & \colhead{interval, eV} & \colhead{$\Delta E_{\rm grain}$, eV} & \colhead{$f_T$, s$^{-1}$, High} & \colhead{$f_T$, s$^{-1}$, Low}
}
\startdata
1 & 10.41 & 10.01--12.00 & 2.00 & 1.30E+02 & 5.530--7.055E+02 & 7.000E+02 & 1.15E-08 & 2.65E-09 \\
2 & 12.73 & 12.01--14.00 & 2.00 & 1.05E+03 & ...--1.742E+03 & 1.037E+03 & 6.57E-10 & 2.18E-11 \\
3 & 14.78 & 14.01--16.00 & 2.00 & 2.26E+03 & ...--3.166E+03 & 1.424E+03 & 1.48E-10 & 5.04E-12 \\
4 & 16.81 & 16.01--18.00 & 2.00 & 3.88E+03 & ...--5.007E+03 & 1.841E+03 & 4.77E-11 & 1.84E-12 \\
5 & 18.87 & 18.01--20.00 & 2.00 & 5.98E+03 & ...--7.315E+03 & 2.308E+03 & 1.91E-11 & 8.52E-13 \\
6 & 20.90 & 20.01--22.00 & 2.00 & 8.55E+03 & ...--1.010E+04 & 2.788E+03 & 8.88E-12 & 4.59E-13 \\
7 & 22.95 & 22.01--24.00 & 2.00 & 1.16E+04 & ...--1.339E+04 & 3.282E+03 & 4.61E-12 & 2.72E-13 \\
8 & 24.92 & 24.01--26.00 & 2.00 & 1.51E+04 & ...--1.717E+04 & 3.783E+03 & 2.75E-12 & 1.86E-13 \\
\enddata
\end{deluxetable}

%
\startlongtable
\begin{deluxetable}{lcccccccc}
\tabletypesize{\footnotesize}
\tablecaption{
Grains with $a=	0.05	$~$\mu$m and 0.025	$\mu$m thick icy mantle,  shielded by interstellar gas with $N_H=	1.98\times10^{22}$ H\,atoms\,cm$^{-2}$.
}
\label{tab4_0.05_9.0}
\tablehead{
\colhead{No.} & \colhead{$T_{\rm CR}$, K} & \colhead{interval, K} & \colhead{$\Delta T_{\rm CR}$, K} & \colhead{$E_{\rm grain}$, eV} & \colhead{interval, eV} & \colhead{$\Delta E_{\rm grain}$, eV} & \colhead{$f_T$, s$^{-1}$, High} & \colhead{$f_T$, s$^{-1}$, Low}
}
\startdata
1 & 10.29 & 10.01--12.00 & 2.00 & 1.55E+02 & 9.176--1.162E+03 & 1.153E+03 & 1.30E-08 & 2.80E-09 \\
2 & 12.69 & 12.01--14.00 & 2.00 & 1.70E+03 & ...--2.844E+03 & 1.681E+03 & 3.18E-10 & 1.23E-11 \\
3 & 14.79 & 14.01--16.00 & 2.00 & 3.69E+03 & ...--5.119E+03 & 2.275E+03 & 5.93E-11 & 2.77E-12 \\
4 & 16.84 & 16.01--18.00 & 2.00 & 6.29E+03 & ...--8.025E+03 & 2.906E+03 & 1.89E-11 & 1.06E-12 \\
5 & 18.87 & 18.01--20.00 & 2.00 & 9.55E+03 & ...--1.163E+04 & 3.606E+03 & 7.86E-12 & 5.26E-13 \\
6 & 20.90 & 20.01--22.00 & 2.00 & 1.35E+04 & ...--1.595E+04 & 4.321E+03 & 3.92E-12 & 2.88E-13 \\
7 & 22.93 & 22.01--24.00 & 2.00 & 1.83E+04 & ...--2.101E+04 & 5.055E+03 & 2.29E-12 & 1.82E-13 \\
8 & 24.97 & 24.01--26.00 & 2.00 & 2.38E+04 & ...--2.680E+04 & 5.798E+03 & 1.35E-12 & 1.17E-13 \\
\enddata
\end{deluxetable}

%
\startlongtable
\begin{deluxetable}{lcccccccc}
\tabletypesize{\footnotesize}
\tablecaption{
Grains with $a=	0.05	$~$\mu$m and 0.03	$\mu$m thick icy mantle,  shielded by interstellar gas with $N_H=	4.40\times10^{22}$ H\,atoms\,cm$^{-2}$.
}
\label{tab5_0.05_20}
\tablehead{
\colhead{No.} & \colhead{$T_{\rm CR}$, K} & \colhead{interval, K} & \colhead{$\Delta T_{\rm CR}$, K} & \colhead{$E_{\rm grain}$, eV} & \colhead{interval, eV} & \colhead{$\Delta E_{\rm grain}$, eV} & \colhead{$f_T$, s$^{-1}$, High} & \colhead{$f_T$, s$^{-1}$, Low}
}
\startdata
1 & 10.21 & 10.01--12.00 & 2.00 & 1.55E+02 & 12.842--1.622E+03 & 1.609E+03 & 1.34E-08 & 2.99E-09 \\
2 & 12.67 & 12.01--14.00 & 2.00 & 2.34E+03 & ...--3.951E+03 & 2.329E+03 & 1.38E-10 & 7.97E-12 \\
3 & 14.80 & 14.01--16.00 & 2.00 & 5.14E+03 & ...--7.080E+03 & 3.130E+03 & 2.45E-11 & 1.88E-12 \\
4 & 16.86 & 16.01--18.00 & 2.00 & 8.71E+03 & ...--1.105E+04 & 3.973E+03 & 8.21E-12 & 7.35E-13 \\
5 & 18.90 & 18.01--20.00 & 2.00 & 1.32E+04 & ...--1.596E+04 & 4.905E+03 & 3.75E-12 & 3.79E-13 \\
6 & 20.88 & 20.01--22.00 & 2.00 & 1.85E+04 & ...--2.181E+04 & 5.855E+03 & 1.87E-12 & 2.04E-13 \\
7 & 22.89 & 22.01--24.00 & 2.00 & 2.48E+04 & ...--2.864E+04 & 6.826E+03 & 1.19E-12 & 1.35E-13 \\
8 & 24.92 & 24.01--26.00 & 2.00 & 3.22E+04 & ...--3.645E+04 & 7.810E+03 & 6.47E-13 & 7.42E-14 \\
\enddata
\end{deluxetable}

%
\startlongtable
\begin{deluxetable}{lcccccccc}
\tabletypesize{\footnotesize}
\tablecaption{
Grains with $a=	0.05	$~$\mu$m and 0.03	$\mu$m thick icy mantle,  shielded by interstellar gas with $N_H=	8.80\times10^{22}$ H\,atoms\,cm$^{-2}$.
}
\label{tab6_0.05_40}
\tablehead{
\colhead{No.} & \colhead{$T_{\rm CR}$, K} & \colhead{interval, K} & \colhead{$\Delta T_{\rm CR}$, K} & \colhead{$E_{\rm grain}$, eV} & \colhead{interval, eV} & \colhead{$\Delta E_{\rm grain}$, eV} & \colhead{$f_T$, s$^{-1}$, High} & \colhead{$f_T$, s$^{-1}$, Low}
}
\startdata
1 & 10.18 & 10.01--12.00 & 2.00 & 1.35E+02 & 12.842--1.622E+03 & 1.609E+03 & 1.23E-08 & 2.95E-09 \\
2 & 12.67 & 12.01--14.00 & 2.00 & 2.34E+03 & ...--3.951E+03 & 2.329E+03 & 8.74E-11 & 6.97E-12 \\
3 & 14.81 & 14.01--16.00 & 2.00 & 5.15E+03 & ...--7.080E+03 & 3.130E+03 & 1.58E-11 & 1.69E-12 \\
4 & 16.86 & 16.01--18.00 & 2.00 & 8.72E+03 & ...--1.105E+04 & 3.973E+03 & 5.51E-12 & 6.57E-13 \\
5 & 18.90 & 18.01--20.00 & 2.00 & 1.32E+04 & ...--1.596E+04 & 4.905E+03 & 2.55E-12 & 3.38E-13 \\
6 & 20.88 & 20.01--22.00 & 2.00 & 1.85E+04 & ...--2.181E+04 & 5.855E+03 & 1.27E-12 & 1.81E-13 \\
7 & 22.89 & 22.01--24.00 & 2.00 & 2.48E+04 & ...--2.864E+04 & 6.826E+03 & 7.99E-13 & 1.19E-13 \\
8 & 24.92 & 24.01--26.00 & 2.00 & 3.22E+04 & ...--3.645E+04 & 7.810E+03 & 4.33E-13 & 6.42E-14 \\
\enddata
\end{deluxetable}

%
\startlongtable
\begin{deluxetable}{lcccccccc}
\tabletypesize{\footnotesize}
\tablecaption{
Grains with $a=	0.05	$~$\mu$m and 0.03	$\mu$m thick icy mantle,  shielded by interstellar gas with $N_H=	1.76\times10^{23}$ H\,atoms\,cm$^{-2}$.
}
\label{tab7_0.05_80}
\tablehead{
\colhead{No.} & \colhead{$T_{\rm CR}$, K} & \colhead{interval, K} & \colhead{$\Delta T_{\rm CR}$, K} & \colhead{$E_{\rm grain}$, eV} & \colhead{interval, eV} & \colhead{$\Delta E_{\rm grain}$, eV} & \colhead{$f_T$, s$^{-1}$, High} & \colhead{$f_T$, s$^{-1}$, Low}
}
\startdata
1 & 10.16 & 10.01--12.00 & 2.00 & 1.17E+02 & 12.842--1.622E+03 & 1.609E+03 & 1.10E-08 & 2.89E-09 \\
2 & 12.67 & 12.01--14.00 & 2.00 & 2.34E+03 & ...--3.951E+03 & 2.329E+03 & 5.45E-11 & 6.03E-12 \\
3 & 14.81 & 14.01--16.00 & 2.00 & 5.16E+03 & ...--7.080E+03 & 3.130E+03 & 1.03E-11 & 1.48E-12 \\
4 & 16.86 & 16.01--18.00 & 2.00 & 8.72E+03 & ...--1.105E+04 & 3.973E+03 & 3.65E-12 & 5.71E-13 \\
5 & 18.89 & 18.01--20.00 & 2.00 & 1.32E+04 & ...--1.596E+04 & 4.905E+03 & 1.68E-12 & 2.91E-13 \\
6 & 20.88 & 20.01--22.00 & 2.00 & 1.85E+04 & ...--2.181E+04 & 5.855E+03 & 8.35E-13 & 1.55E-13 \\
7 & 22.89 & 22.01--24.00 & 2.00 & 2.48E+04 & ...--2.864E+04 & 6.826E+03 & 5.15E-13 & 1.01E-13 \\
8 & 24.92 & 24.01--26.00 & 2.00 & 3.22E+04 & ...--3.645E+04 & 7.810E+03 & 2.78E-13 & 5.33E-14 \\
\enddata
\end{deluxetable}

%
\startlongtable
\begin{deluxetable}{lcccccccc}
\tabletypesize{\footnotesize}
\tablecaption{
Grains with $a=	0.05	$~$\mu$m and 0.03	$\mu$m thick icy mantle,  shielded by interstellar gas with $N_H=	3.52\times10^{23}$ H\,atoms\,cm$^{-2}$.
}
\label{tab8_0.05_160}
\tablehead{
\colhead{No.} & \colhead{$T_{\rm CR}$, K} & \colhead{interval, K} & \colhead{$\Delta T_{\rm CR}$, K} & \colhead{$E_{\rm grain}$, eV} & \colhead{interval, eV} & \colhead{$\Delta E_{\rm grain}$, eV} & \colhead{$f_T$, s$^{-1}$, High} & \colhead{$f_T$, s$^{-1}$, Low}
}
\startdata
1 & 10.13 & 10.01--12.00 & 2.00 & 1.01E+02 & 12.842--1.622E+03 & 1.609E+03 & 9.75E-09 & 2.81E-09 \\
2 & 12.68 & 12.01--14.00 & 2.00 & 2.35E+03 & ...--3.951E+03 & 2.329E+03 & 3.37E-11 & 5.14E-12 \\
3 & 14.82 & 14.01--16.00 & 2.00 & 5.16E+03 & ...--7.080E+03 & 3.130E+03 & 6.67E-12 & 1.27E-12 \\
4 & 16.85 & 16.01--18.00 & 2.00 & 8.71E+03 & ...--1.105E+04 & 3.973E+03 & 2.36E-12 & 4.78E-13 \\
5 & 18.89 & 18.01--20.00 & 2.00 & 1.32E+04 & ...--1.596E+04 & 4.905E+03 & 1.07E-12 & 2.42E-13 \\
6 & 20.88 & 20.01--22.00 & 2.00 & 1.85E+04 & ...--2.181E+04 & 5.855E+03 & 5.29E-13 & 1.27E-13 \\
7 & 22.89 & 22.01--24.00 & 2.00 & 2.48E+04 & ...--2.864E+04 & 6.826E+03 & 3.23E-13 & 8.29E-14 \\
8 & 24.92 & 24.01--26.00 & 2.00 & 3.22E+04 & ...--3.645E+04 & 7.810E+03 & 1.72E-13 & 4.23E-14 \\
\enddata
\end{deluxetable}

%
\startlongtable
\begin{deluxetable}{lcccccccc}
\tabletypesize{\footnotesize}
\tablecaption{
Bare grains with $a=0.1$~$\mu$m, shielded by interstellar gas with $N_H=3.30\times10^{21}$ H\,atoms\,cm$^{-2}$.
}
\label{tab9_0.1_1.5}
\tablehead{
\colhead{No.} & \colhead{$T_{\rm CR}$, K} & \colhead{interval, K} & \colhead{$\Delta T_{\rm CR}$, K} & \colhead{$E_{\rm grain}$, eV} & \colhead{interval, eV} & \colhead{$\Delta E_{\rm grain}$, eV} & \colhead{$f_T$, s$^{-1}$, High} & \colhead{$f_T$, s$^{-1}$, Low}
}
\startdata
1 & 10.60 & 10.01--12.00 & 2.00 & 2.01E+02 & 6.3238--6.752E+02 & 6.689E+02 & 2.05E-08 & 5.25E-09 \\
2 & 12.85 & 12.01--14.00 & 2.00 & 9.97E+02 & ...--1.465E+03 & 7.900E+02 & 3.83E-09 & 1.34E-10 \\
3 & 14.88 & 14.01--16.00 & 2.00 & 1.88E+03 & ...--2.438E+03 & 9.729E+02 & 1.73E-09 & 4.00E-11 \\
4 & 16.90 & 16.01--18.00 & 2.00 & 2.96E+03 & ...--3.648E+03 & 1.210E+03 & 8.99E-10 & 1.74E-11 \\
5 & 18.90 & 18.01--20.00 & 2.00 & 4.31E+03 & ...--5.164E+03 & 1.516E+03 & 4.88E-10 & 9.16E-12 \\
6 & 20.90 & 20.01--22.00 & 2.00 & 5.98E+03 & ...--7.037E+03 & 1.873E+03 & 2.53E-10 & 5.08E-12 \\
7 & 22.89 & 22.01--24.00 & 2.00 & 8.03E+03 & ...--9.319E+03 & 2.282E+03 & 1.62E-10 & 3.18E-12 \\
8 & 24.99 & 24.01--26.00 & 2.00 & 1.07E+04 & ...--1.206E+04 & 2.741E+03 & 8.72E-11 & 1.89E-12 \\
\enddata
\end{deluxetable}

%
\startlongtable
\begin{deluxetable}{lcccccccc}
\tabletypesize{\footnotesize}
\tablecaption{
Grains with $a=	0.1	$~$\mu$m and 0.01	$\mu$m thick icy mantle,  shielded by interstellar gas with $N_H=	6.60\times10^{21}$ H\,atoms\,cm$^{-2}$.
}
\label{tab10_0.1_3.0}
\tablehead{
\colhead{No.} & \colhead{$T_{\rm CR}$, K} & \colhead{interval, K} & \colhead{$\Delta T_{\rm CR}$, K} & \colhead{$E_{\rm grain}$, eV} & \colhead{interval, eV} & \colhead{$\Delta E_{\rm grain}$, eV} & \colhead{$f_T$, s$^{-1}$, High} & \colhead{$f_T$, s$^{-1}$, Low}
}
\startdata
1 & 10.46 & 10.01--12.00 & 2.00 & 2.81E+02 & 10.987--1.310E+03 & 1.299E+03 & 2.87E-08 & 6.52E-09 \\
2 & 12.76 & 12.01--14.00 & 2.00 & 1.94E+03 & ...--3.107E+03 & 1.797E+03 & 2.56E-09 & 7.01E-11 \\
3 & 14.80 & 14.01--16.00 & 2.00 & 4.02E+03 & ...--5.525E+03 & 2.418E+03 & 7.12E-10 & 1.75E-11 \\
4 & 16.85 & 16.01--18.00 & 2.00 & 6.80E+03 & ...--8.654E+03 & 3.129E+03 & 2.48E-10 & 6.52E-12 \\
5 & 18.88 & 18.01--20.00 & 2.00 & 1.03E+04 & ...--1.261E+04 & 3.960E+03 & 1.02E-10 & 3.12E-12 \\
6 & 20.88 & 20.01--22.00 & 2.00 & 1.47E+04 & ...--1.747E+04 & 4.852E+03 & 5.01E-11 & 1.62E-12 \\
7 & 22.92 & 22.01--24.00 & 2.00 & 2.01E+04 & ...--2.327E+04 & 5.804E+03 & 2.25E-11 & 9.60E-13 \\
8 & 24.87 & 24.01--26.00 & 2.00 & 2.62E+04 & ...--3.008E+04 & 6.807E+03 & 1.59E-11 & 6.65E-13 \\
\enddata
\end{deluxetable}

%
\startlongtable
\begin{deluxetable}{lcccccccc}
\tabletypesize{\footnotesize}
\tablecaption{
Grains with $a=	0.1	$~$\mu$m and 0.020	$\mu$m thick icy mantle,  shielded by interstellar gas with $N_H=	1.32\times10^{22}$ H\,atoms\,cm$^{-2}$.
}
\label{tab11_0.1_6.0}
\tablehead{
\colhead{No.} & \colhead{$T_{\rm CR}$, K} & \colhead{interval, K} & \colhead{$\Delta T_{\rm CR}$, K} & \colhead{$E_{\rm grain}$, eV} & \colhead{interval, eV} & \colhead{$\Delta E_{\rm grain}$, eV} & \colhead{$f_T$, s$^{-1}$, High} & \colhead{$f_T$, s$^{-1}$, Low}
}
\startdata
1 & 10.22 & 10.01--11.00 & 1.00 & 2.61E+02 & 21.603--1.215E+03 & 1.193E+03 & 3.11E-08 & 6.62E-09 \\
2 & 11.38 & 11.01--12.00 & 1.00 & 1.74E+03 & ...--2.669E+03 & 1.454E+03 & 2.32E-09 & 7.88E-11 \\
3 & 12.42 & 12.01--13.00 & 1.00 & 3.37E+03 & ...--4.408E+03 & 1.739E+03 & 6.29E-10 & 2.02E-11 \\
4 & 13.43 & 13.01--14.00 & 1.00 & 5.26E+03 & ...--6.457E+03 & 2.050E+03 & 2.46E-10 & 8.45E-12 \\
5 & 14.45 & 14.01--15.00 & 1.00 & 7.50E+03 & ...--8.843E+03 & 2.385E+03 & 1.17E-10 & 4.28E-12 \\
6 & 15.47 & 15.01--16.00 & 1.00 & 1.01E+04 & ...--1.159E+04 & 2.743E+03 & 5.66E-11 & 2.38E-12 \\
7 & 16.49 & 16.01--17.00 & 1.00 & 1.31E+04 & ...--1.467E+04 & 3.088E+03 & 3.48E-11 & 1.61E-12 \\
8 & 17.49 & 17.01--18.00 & 1.00 & 1.64E+04 & ...--1.819E+04 & 3.513E+03 & 2.04E-11 & 9.86E-13 \\
\enddata
\end{deluxetable}

%
\startlongtable
\begin{deluxetable}{lcccccccc}
\tabletypesize{\footnotesize}
\tablecaption{
Grains with $a=	0.1	$~$\mu$m and 0.025	$\mu$m thick icy mantle,  shielded by interstellar gas with $N_H=	1.98\times10^{22}$ H\,atoms\,cm$^{-2}$.
}
\label{tab12_0.1_9.0}
\tablehead{
\colhead{No.} & \colhead{$T_{\rm CR}$, K} & \colhead{interval, K} & \colhead{$\Delta T_{\rm CR}$, K} & \colhead{$E_{\rm grain}$, eV} & \colhead{interval, eV} & \colhead{$\Delta E_{\rm grain}$, eV} & \colhead{$f_T$, s$^{-1}$, High} & \colhead{$f_T$, s$^{-1}$, Low}
}
\startdata
1 & 10.17 & 10.01--11.00 & 1.00 & 3.17E+02 & 33.247--1.878E+03 & 1.845E+03 & 3.23E-08 & 6.48E-09 \\
2 & 11.36 & 11.01--12.00 & 1.00 & 2.65E+03 & ...--4.136E+03 & 2.258E+03 & 1.22E-09 & 4.58E-11 \\
3 & 12.41 & 12.01--13.00 & 1.00 & 5.20E+03 & ...--6.834E+03 & 2.699E+03 & 2.85E-10 & 1.14E-11 \\
4 & 13.42 & 13.01--14.00 & 1.00 & 8.14E+03 & ...--1.001E+04 & 3.173E+03 & 9.92E-11 & 4.70E-12 \\
5 & 14.44 & 14.01--15.00 & 1.00 & 1.16E+04 & ...--1.368E+04 & 3.677E+03 & 4.68E-11 & 2.46E-12 \\
6 & 15.46 & 15.01--16.00 & 1.00 & 1.56E+04 & ...--1.789E+04 & 4.209E+03 & 2.42E-11 & 1.38E-12 \\
7 & 16.44 & 16.01--17.00 & 1.00 & 2.00E+04 & ...--2.261E+04 & 4.715E+03 & 1.37E-11 & 8.62E-13 \\
8 & 17.47 & 17.01--18.00 & 1.00 & 2.51E+04 & ...--2.795E+04 & 5.338E+03 & 8.95E-12 & 6.30E-13 \\
\enddata
\end{deluxetable}

%
\startlongtable
\begin{deluxetable}{lcccccccc}
\tabletypesize{\footnotesize}
\tablecaption{
Grains with $a=	0.1	$~$\mu$m and 0.03	$\mu$m thick icy mantle,  shielded by interstellar gas with $N_H=	4.40\times10^{22}$ H\,atoms\,cm$^{-2}$.
}
\label{tab13_0.1_20}
\tablehead{
\colhead{No.} & \colhead{$T_{\rm CR}$, K} & \colhead{interval, K} & \colhead{$\Delta T_{\rm CR}$, K} & \colhead{$E_{\rm grain}$, eV} & \colhead{interval, eV} & \colhead{$\Delta E_{\rm grain}$, eV} & \colhead{$f_T$, s$^{-1}$, High} & \colhead{$f_T$, s$^{-1}$, Low}
}
\startdata
1 & 10.13 & 10.01--11.00 & 1.00 & 3.35E+02 & 45.014--2.553E+03 & 2.508E+03 & 3.01E-08 & 5.91E-09 \\
2 & 11.36 & 11.01--12.00 & 1.00 & 3.61E+03 & ...--5.636E+03 & 3.083E+03 & 5.22E-10 & 2.84E-11 \\
3 & 12.40 & 12.01--13.00 & 1.00 & 7.05E+03 & ...--9.329E+03 & 3.693E+03 & 1.14E-10 & 7.24E-12 \\
4 & 13.43 & 13.01--14.00 & 1.00 & 1.11E+04 & ...--1.367E+04 & 4.342E+03 & 3.99E-11 & 3.10E-12 \\
5 & 14.46 & 14.01--15.00 & 1.00 & 1.59E+04 & ...--1.870E+04 & 5.027E+03 & 1.84E-11 & 1.51E-12 \\
6 & 15.45 & 15.01--16.00 & 1.00 & 2.12E+04 & ...--2.444E+04 & 5.745E+03 & 1.03E-11 & 9.14E-13 \\
7 & 16.48 & 16.01--17.00 & 1.00 & 2.75E+04 & ...--3.087E+04 & 6.424E+03 & 6.07E-12 & 6.07E-13 \\
8 & 17.48 & 17.01--18.00 & 1.00 & 3.44E+04 & ...--3.812E+04 & 7.257E+03 & 4.04E-12 & 4.32E-13 \\
\enddata
\end{deluxetable}

%
\startlongtable
\begin{deluxetable}{lcccccccc}
\tabletypesize{\footnotesize}
\tablecaption{
Grains with $a=	0.1	$~$\mu$m and 0.03	$\mu$m thick icy mantle,  shielded by interstellar gas with $N_H=	8.80\times10^{22}$ H\,atoms\,cm$^{-2}$.
}
\label{tab14_0.1_40}
\tablehead{
\colhead{No.} & \colhead{$T_{\rm CR}$, K} & \colhead{interval, K} & \colhead{$\Delta T_{\rm CR}$, K} & \colhead{$E_{\rm grain}$, eV} & \colhead{interval, eV} & \colhead{$\Delta E_{\rm grain}$, eV} & \colhead{$f_T$, s$^{-1}$, High} & \colhead{$f_T$, s$^{-1}$, Low}
}
\startdata
1 & 10.11 & 10.01--11.00 & 1.00 & 2.99E+02 & 45.014--2.553E+03 & 2.508E+03 & 2.72E-08 & 5.82E-09 \\
2 & 11.36 & 11.01--12.00 & 1.00 & 3.60E+03 & ...--5.636E+03 & 3.083E+03 & 3.35E-10 & 2.47E-11 \\
3 & 12.40 & 12.01--13.00 & 1.00 & 7.06E+03 & ...--9.329E+03 & 3.693E+03 & 7.12E-11 & 6.39E-12 \\
4 & 13.44 & 13.01--14.00 & 1.00 & 1.12E+04 & ...--1.367E+04 & 4.342E+03 & 2.58E-11 & 2.78E-12 \\
5 & 14.46 & 14.01--15.00 & 1.00 & 1.59E+04 & ...--1.870E+04 & 5.027E+03 & 1.21E-11 & 1.34E-12 \\
6 & 15.45 & 15.01--16.00 & 1.00 & 2.12E+04 & ...--2.444E+04 & 5.745E+03 & 7.00E-12 & 8.09E-13 \\
7 & 16.47 & 16.01--17.00 & 1.00 & 2.74E+04 & ...--3.087E+04 & 6.424E+03 & 4.12E-12 & 5.40E-13 \\
8 & 17.48 & 17.01--18.00 & 1.00 & 3.43E+04 & ...--3.812E+04 & 7.257E+03 & 2.77E-12 & 3.83E-13 \\
\enddata
\end{deluxetable}

%
\startlongtable
\begin{deluxetable}{lcccccccc}
\tabletypesize{\footnotesize}
\tablecaption{
Grains with $a=	0.1	$~$\mu$m and 0.03	$\mu$m thick icy mantle,  shielded by interstellar gas with $N_H=	1.76\times10^{23}$ H\,atoms\,cm$^{-2}$.
}
\label{tab15_0.1_80}
\tablehead{
\colhead{No.} & \colhead{$T_{\rm CR}$, K} & \colhead{interval, K} & \colhead{$\Delta T_{\rm CR}$, K} & \colhead{$E_{\rm grain}$, eV} & \colhead{interval, eV} & \colhead{$\Delta E_{\rm grain}$, eV} & \colhead{$f_T$, s$^{-1}$, High} & \colhead{$f_T$, s$^{-1}$, Low}
}
\startdata
1 & 10.10 & 10.01--11.00 & 1.00 & 2.64E+02 & 45.014--2.553E+03 & 2.508E+03 & 2.41E-08 & 5.68E-09 \\
2 & 11.35 & 11.01--12.00 & 1.00 & 3.60E+03 & ...--5.636E+03 & 3.083E+03 & 2.09E-10 & 2.12E-11 \\
3 & 12.40 & 12.01--13.00 & 1.00 & 7.08E+03 & ...--9.329E+03 & 3.693E+03 & 4.44E-11 & 5.58E-12 \\
4 & 13.44 & 13.01--14.00 & 1.00 & 1.12E+04 & ...--1.367E+04 & 4.342E+03 & 1.68E-11 & 2.45E-12 \\
5 & 14.45 & 14.01--15.00 & 1.00 & 1.59E+04 & ...--1.870E+04 & 5.027E+03 & 7.94E-12 & 1.17E-12 \\
6 & 15.44 & 15.01--16.00 & 1.00 & 2.12E+04 & ...--2.444E+04 & 5.745E+03 & 4.64E-12 & 6.93E-13 \\
7 & 16.47 & 16.01--17.00 & 1.00 & 2.74E+04 & ...--3.087E+04 & 6.424E+03 & 2.71E-12 & 4.66E-13 \\
8 & 17.47 & 17.01--18.00 & 1.00 & 3.43E+04 & ...--3.812E+04 & 7.257E+03 & 1.83E-12 & 3.29E-13 \\
\enddata
\end{deluxetable}

%
\startlongtable
\begin{deluxetable}{lcccccccc}
\tabletypesize{\footnotesize}
\tablecaption{
Grains with $a=	0.1	$~$\mu$m and 0.03	$\mu$m thick icy mantle,  shielded by interstellar gas with $N_H=	3.52\times10^{23}$ H\,atoms\,cm$^{-2}$.
}
\label{tab16_0.1_160}
\tablehead{
\colhead{No.} & \colhead{$T_{\rm CR}$, K} & \colhead{interval, K} & \colhead{$\Delta T_{\rm CR}$, K} & \colhead{$E_{\rm grain}$, eV} & \colhead{interval, eV} & \colhead{$\Delta E_{\rm grain}$, eV} & \colhead{$f_T$, s$^{-1}$, High} & \colhead{$f_T$, s$^{-1}$, Low}
}
\startdata
1 & 10.09 & 10.01--11.00 & 1.00 & 2.33E+02 & 45.014--2.553E+03 & 2.508E+03 & 2.09E-08 & 5.48E-09 \\
2 & 11.35 & 11.01--12.00 & 1.00 & 3.60E+03 & ...--5.636E+03 & 3.083E+03 & 1.28E-10 & 1.80E-11 \\
3 & 12.41 & 12.01--13.00 & 1.00 & 7.10E+03 & ...--9.329E+03 & 3.693E+03 & 2.79E-11 & 4.80E-12 \\
4 & 13.44 & 13.01--14.00 & 1.00 & 1.12E+04 & ...--1.367E+04 & 4.342E+03 & 1.09E-11 & 2.10E-12 \\
5 & 14.45 & 14.01--15.00 & 1.00 & 1.59E+04 & ...--1.870E+04 & 5.027E+03 & 5.11E-12 & 9.82E-13 \\
6 & 15.44 & 15.01--16.00 & 1.00 & 2.11E+04 & ...--2.444E+04 & 5.745E+03 & 2.96E-12 & 5.67E-13 \\
7 & 16.47 & 16.01--17.00 & 1.00 & 2.74E+04 & ...--3.087E+04 & 6.424E+03 & 1.72E-12 & 3.87E-13 \\
8 & 17.47 & 17.01--18.00 & 1.00 & 3.43E+04 & ...--3.812E+04 & 7.257E+03 & 1.16E-12 & 2.72E-13 \\
\enddata
\end{deluxetable}

%
\startlongtable
\begin{deluxetable}{lcccccccc}
\tabletypesize{\footnotesize}
\tablecaption{
Bare grains with $a=0.2$~$\mu$m, shielded by interstellar gas with $N_H=3.30\times10^{21}$ H\,atoms\,cm$^{-2}$.
}
\label{tab17_0.2_1.5}
\tablehead{
\colhead{No.} & \colhead{$T_{\rm CR}$, K} & \colhead{interval, K} & \colhead{$\Delta T_{\rm CR}$, K} & \colhead{$E_{\rm grain}$, eV} & \colhead{interval, eV} & \colhead{$\Delta E_{\rm grain}$, eV} & \colhead{$f_T$, s$^{-1}$, High} & \colhead{$f_T$, s$^{-1}$, Low}
}
\startdata
1 & 10.23 & 10.01--11.00 & 1.00 & 6.24E+02 & 50.5902--2.633E+03 & 2.582E+03 & 9.42E-08 & 2.14E-08 \\
2 & 11.40 & 11.01--12.00 & 1.00 & 3.74E+03 & ...--5.402E+03 & 2.769E+03 & 1.05E-08 & 2.57E-10 \\
3 & 12.44 & 12.01--13.00 & 1.00 & 6.70E+03 & ...--8.410E+03 & 3.008E+03 & 3.60E-09 & 6.91E-11 \\
4 & 13.45 & 13.01--14.00 & 1.00 & 9.86E+03 & ...--1.172E+04 & 3.312E+03 & 1.74E-09 & 3.28E-11 \\
5 & 14.44 & 14.01--15.00 & 1.00 & 1.33E+04 & ...--1.540E+04 & 3.678E+03 & 8.07E-10 & 1.59E-11 \\
6 & 15.39 & 15.01--16.00 & 1.00 & 1.70E+04 & ...--1.950E+04 & 4.105E+03 & 5.48E-10 & 1.05E-11 \\
7 & 16.44 & 16.01--17.00 & 1.00 & 2.15E+04 & ...--2.405E+04 & 4.544E+03 & 3.37E-10 & 7.08E-12 \\
8 & 17.48 & 17.01--18.00 & 1.00 & 2.65E+04 & ...--2.918E+04 & 5.133E+03 & 2.23E-10 & 4.60E-12 \\
\enddata
\end{deluxetable}

%
\startlongtable
\begin{deluxetable}{lcccccccc}
\tabletypesize{\footnotesize}
\tablecaption{
Grains with $a=	0.2	$~$\mu$m and 0.01	$\mu$m thick icy mantle,  shielded by interstellar gas with $N_H=	6.60\times10^{21}$ H\,atoms\,cm$^{-2}$.
}
\label{tab18_0.2_3.0}
\tablehead{
\colhead{No.} & \colhead{$T_{\rm CR}$, K} & \colhead{interval, K} & \colhead{$\Delta T_{\rm CR}$, K} & \colhead{$E_{\rm grain}$, eV} & \colhead{interval, eV} & \colhead{$\Delta E_{\rm grain}$, eV} & \colhead{$f_T$, s$^{-1}$, High} & \colhead{$f_T$, s$^{-1}$, Low}
}
\startdata
1 & 10.20 & 10.01--11.00 & 1.00 & 7.44E+02 & 68.357--3.686E+03 & 3.617E+03 & 1.02E-07 & 2.01E-08 \\
2 & 11.38 & 11.01--12.00 & 1.00 & 5.20E+03 & ...--7.819E+03 & 4.133E+03 & 6.67E-09 & 1.73E-10 \\
3 & 12.41 & 12.01--13.00 & 1.00 & 9.72E+03 & ...--1.255E+04 & 4.732E+03 & 1.77E-09 & 4.47E-11 \\
4 & 13.43 & 13.01--14.00 & 1.00 & 1.48E+04 & ...--1.798E+04 & 5.426E+03 & 7.40E-10 & 1.95E-11 \\
5 & 14.47 & 14.01--15.00 & 1.00 & 2.08E+04 & ...--2.419E+04 & 6.210E+03 & 3.33E-10 & 9.56E-12 \\
6 & 15.49 & 15.01--16.00 & 1.00 & 2.76E+04 & ...--3.127E+04 & 7.079E+03 & 1.91E-10 & 5.74E-12 \\
7 & 16.49 & 16.01--17.00 & 1.00 & 3.51E+04 & ...--3.921E+04 & 7.944E+03 & 9.65E-11 & 3.63E-12 \\
8 & 17.46 & 17.01--18.00 & 1.00 & 4.33E+04 & ...--4.825E+04 & 9.044E+03 & 6.13E-11 & 2.69E-12 \\
\enddata
\end{deluxetable}

%
\startlongtable
\begin{deluxetable}{lcccccccc}
\tabletypesize{\footnotesize}
\tablecaption{
Grains with $a=	0.2	$~$\mu$m and 0.020	$\mu$m thick icy mantle,  shielded by interstellar gas with $N_H=	1.32\times10^{22}$ H\,atoms\,cm$^{-2}$.
}
\label{tab19_0.2_6.0}
\tablehead{
\colhead{No.} & \colhead{$T_{\rm CR}$, K} & \colhead{interval, K} & \colhead{$\Delta T_{\rm CR}$, K} & \colhead{$E_{\rm grain}$, eV} & \colhead{interval, eV} & \colhead{$\Delta E_{\rm grain}$, eV} & \colhead{$f_T$, s$^{-1}$, High} & \colhead{$f_T$, s$^{-1}$, Low}
}
\startdata
1 & 10.15 & 10.01--11.00 & 1.00 & 9.06E+02 & 106.165--5.855E+03 & 5.749E+03 & 9.93E-08 & 1.65E-08 \\
2 & 11.36 & 11.01--12.00 & 1.00 & 8.23E+03 & ...--1.265E+04 & 6.800E+03 & 2.65E-09 & 8.65E-11 \\
3 & 12.40 & 12.01--13.00 & 1.00 & 1.58E+04 & ...--2.062E+04 & 7.964E+03 & 6.38E-10 & 2.36E-11 \\
4 & 13.45 & 13.01--14.00 & 1.00 & 2.47E+04 & ...--2.988E+04 & 9.261E+03 & 2.15E-10 & 9.30E-12 \\
5 & 14.43 & 14.01--15.00 & 1.00 & 3.43E+04 & ...--4.056E+04 & 1.068E+04 & 9.99E-11 & 4.78E-12 \\
6 & 15.44 & 15.01--16.00 & 1.00 & 4.58E+04 & ...--5.278E+04 & 1.222E+04 & 5.39E-11 & 3.13E-12 \\
7 & 16.42 & 16.01--17.00 & 1.00 & 5.85E+04 & ...--6.649E+04 & 1.371E+04 & 3.43E-11 & 2.09E-12 \\
8 & 17.48 & 17.01--18.00 & 1.00 & 7.39E+04 & ...--8.207E+04 & 1.558E+04 & 2.21E-11 & 1.45E-12 \\
\enddata
\end{deluxetable}

%
\startlongtable
\begin{deluxetable}{lcccccccc}
\tabletypesize{\footnotesize}
\tablecaption{
Grains with $a=	0.2	$~$\mu$m and 0.025	$\mu$m thick icy mantle,  shielded by interstellar gas with $N_H=	1.98\times10^{22}$ H\,atoms\,cm$^{-2}$.
}
\label{tab20_0.2_9.0}
\tablehead{
\colhead{No.} & \colhead{$T_{\rm CR}$, K} & \colhead{interval, K} & \colhead{$\Delta T_{\rm CR}$, K} & \colhead{$E_{\rm grain}$, eV} & \colhead{interval, eV} & \colhead{$\Delta E_{\rm grain}$, eV} & \colhead{$f_T$, s$^{-1}$, High} & \colhead{$f_T$, s$^{-1}$, Low}
}
\startdata
1 & 10.12 & 10.01--11.00 & 1.00 & 1.02E+03 & 146.366--8.144E+03 & 7.997E+03 & 8.98E-08 & 1.26E-08 \\
2 & 11.35 & 11.01--12.00 & 1.00 & 1.14E+04 & ...--1.771E+04 & 9.569E+03 & 1.36E-09 & 5.56E-11 \\
3 & 12.42 & 12.01--13.00 & 1.00 & 2.23E+04 & ...--2.898E+04 & 1.127E+04 & 2.76E-10 & 1.42E-11 \\
4 & 13.43 & 13.01--14.00 & 1.00 & 3.44E+04 & ...--4.211E+04 & 1.313E+04 & 9.78E-11 & 5.78E-12 \\
5 & 14.45 & 14.01--15.00 & 1.00 & 4.87E+04 & ...--5.723E+04 & 1.512E+04 & 5.01E-11 & 3.46E-12 \\
6 & 15.43 & 15.01--16.00 & 1.00 & 6.45E+04 & ...--7.449E+04 & 1.725E+04 & 2.53E-11 & 1.92E-12 \\
7 & 16.44 & 16.01--17.00 & 1.00 & 8.28E+04 & ...--9.378E+04 & 1.930E+04 & 1.83E-11 & 1.49E-12 \\
8 & 17.48 & 17.01--18.00 & 1.00 & 1.04E+05 & ...--1.156E+05 & 2.184E+04 & 1.07E-11 & 8.87E-13 \\
\enddata
\end{deluxetable}

%
\startlongtable
\begin{deluxetable}{lcccccccc}
\tabletypesize{\footnotesize}
\tablecaption{
Grains with $a=	0.2	$~$\mu$m and 0.03	$\mu$m thick icy mantle,  shielded by interstellar gas with $N_H=	4.40\times10^{22}$ H\,atoms\,cm$^{-2}$.
}
\label{tab21_0.2_20}
\tablehead{
\colhead{No.} & \colhead{$T_{\rm CR}$, K} & \colhead{interval, K} & \colhead{$\Delta T_{\rm CR}$, K} & \colhead{$E_{\rm grain}$, eV} & \colhead{interval, eV} & \colhead{$\Delta E_{\rm grain}$, eV} & \colhead{$f_T$, s$^{-1}$, High} & \colhead{$f_T$, s$^{-1}$, Low}
}
\startdata
1 & 10.09 & 10.01--11.00 & 1.00 & 1.01E+03 & 185.277--1.037E+04 & 1.019E+04 & 7.56E-08 & 1.02E-08 \\
2 & 11.35 & 11.01--12.00 & 1.00 & 1.45E+04 & ...--2.267E+04 & 1.230E+04 & 5.54E-10 & 3.58E-11 \\
3 & 12.41 & 12.01--13.00 & 1.00 & 2.85E+04 & ...--3.723E+04 & 1.455E+04 & 1.23E-10 & 9.60E-12 \\
4 & 13.44 & 13.01--14.00 & 1.00 & 4.45E+04 & ...--5.421E+04 & 1.699E+04 & 4.33E-11 & 4.12E-12 \\
5 & 14.43 & 14.01--15.00 & 1.00 & 6.25E+04 & ...--7.379E+04 & 1.958E+04 & 2.33E-11 & 2.31E-12 \\
6 & 15.45 & 15.01--16.00 & 1.00 & 8.37E+04 & ...--9.611E+04 & 2.232E+04 & 1.31E-11 & 1.46E-12 \\
7 & 16.47 & 16.01--17.00 & 1.00 & 1.08E+05 & ...--1.210E+05 & 2.493E+04 & 8.11E-12 & 9.34E-13 \\
8 & 17.48 & 17.01--18.00 & 1.00 & 1.35E+05 & ...--1.492E+05 & 2.816E+04 & 5.19E-12 & 5.60E-13 \\
\enddata
\end{deluxetable}

%
\startlongtable
\begin{deluxetable}{lcccccccc}
\tabletypesize{\footnotesize}
\tablecaption{
Grains with $a=	0.2	$~$\mu$m and 0.03	$\mu$m thick icy mantle,  shielded by interstellar gas with $N_H=	8.80\times10^{22}$ H\,atoms\,cm$^{-2}$.
}
\label{tab22_0.2_40}
\tablehead{
\colhead{No.} & \colhead{$T_{\rm CR}$, K} & \colhead{interval, K} & \colhead{$\Delta T_{\rm CR}$, K} & \colhead{$E_{\rm grain}$, eV} & \colhead{interval, eV} & \colhead{$\Delta E_{\rm grain}$, eV} & \colhead{$f_T$, s$^{-1}$, High} & \colhead{$f_T$, s$^{-1}$, Low}
}
\startdata
1 & 10.08 & 10.01--11.00 & 1.00 & 8.94E+02 & 185.277--1.037E+04 & 1.019E+04 & 6.63E-08 & 9.86E-09 \\
2 & 11.35 & 11.01--12.00 & 1.00 & 1.45E+04 & ...--2.267E+04 & 1.230E+04 & 3.49E-10 & 3.17E-11 \\
3 & 12.42 & 12.01--13.00 & 1.00 & 2.85E+04 & ...--3.723E+04 & 1.455E+04 & 7.98E-11 & 8.55E-12 \\
4 & 13.44 & 13.01--14.00 & 1.00 & 4.45E+04 & ...--5.421E+04 & 1.699E+04 & 2.93E-11 & 3.67E-12 \\
5 & 14.43 & 14.01--15.00 & 1.00 & 6.25E+04 & ...--7.379E+04 & 1.958E+04 & 1.58E-11 & 2.05E-12 \\
6 & 15.45 & 15.01--16.00 & 1.00 & 8.36E+04 & ...--9.611E+04 & 2.232E+04 & 8.87E-12 & 1.30E-12 \\
7 & 16.46 & 16.01--17.00 & 1.00 & 1.08E+05 & ...--1.210E+05 & 2.493E+04 & 5.49E-12 & 8.19E-13 \\
8 & 17.48 & 17.01--18.00 & 1.00 & 1.35E+05 & ...--1.492E+05 & 2.816E+04 & 3.42E-12 & 4.83E-13 \\
\enddata
\end{deluxetable}

%
\startlongtable
\begin{deluxetable}{lcccccccc}
\tabletypesize{\footnotesize}
\tablecaption{
Grains with $a=	0.2	$~$\mu$m and 0.03	$\mu$m thick icy mantle,  shielded by interstellar gas with $N_H=	1.76\times10^{23}$ H\,atoms\,cm$^{-2}$.
}
\label{tab23_0.2_80}
\tablehead{
\colhead{No.} & \colhead{$T_{\rm CR}$, K} & \colhead{interval, K} & \colhead{$\Delta T_{\rm CR}$, K} & \colhead{$E_{\rm grain}$, eV} & \colhead{interval, eV} & \colhead{$\Delta E_{\rm grain}$, eV} & \colhead{$f_T$, s$^{-1}$, High} & \colhead{$f_T$, s$^{-1}$, Low}
}
\startdata
1 & 10.07 & 10.01--11.00 & 1.00 & 7.88E+02 & 185.277--1.037E+04 & 1.019E+04 & 5.67E-08 & 9.46E-09 \\
2 & 11.36 & 11.01--12.00 & 1.00 & 1.46E+04 & ...--2.267E+04 & 1.230E+04 & 2.20E-10 & 2.77E-11 \\
3 & 12.42 & 12.01--13.00 & 1.00 & 2.86E+04 & ...--3.723E+04 & 1.455E+04 & 5.19E-11 & 7.44E-12 \\
4 & 13.44 & 13.01--14.00 & 1.00 & 4.46E+04 & ...--5.421E+04 & 1.699E+04 & 1.94E-11 & 3.18E-12 \\
5 & 14.43 & 14.01--15.00 & 1.00 & 6.25E+04 & ...--7.379E+04 & 1.958E+04 & 1.03E-11 & 1.75E-12 \\
6 & 15.45 & 15.01--16.00 & 1.00 & 8.36E+04 & ...--9.611E+04 & 2.232E+04 & 5.79E-12 & 1.12E-12 \\
7 & 16.46 & 16.01--17.00 & 1.00 & 1.08E+05 & ...--1.210E+05 & 2.493E+04 & 3.57E-12 & 6.94E-13 \\
8 & 17.48 & 17.01--18.00 & 1.00 & 1.35E+05 & ...--1.492E+05 & 2.816E+04 & 2.17E-12 & 4.00E-13 \\
\enddata
\end{deluxetable}

%
\startlongtable
\begin{deluxetable}{lcccccccc}
\tabletypesize{\footnotesize}
\tablecaption{
Grains with $a=	0.2	$~$\mu$m and 0.03	$\mu$m thick icy mantle,  shielded by interstellar gas with $N_H=	3.52\times10^{23}$ H\,atoms\,cm$^{-2}$.
}
\label{tab24_0.2_160}
\tablehead{
\colhead{No.} & \colhead{$T_{\rm CR}$, K} & \colhead{interval, K} & \colhead{$\Delta T_{\rm CR}$, K} & \colhead{$E_{\rm grain}$, eV} & \colhead{interval, eV} & \colhead{$\Delta E_{\rm grain}$, eV} & \colhead{$f_T$, s$^{-1}$, High} & \colhead{$f_T$, s$^{-1}$, Low}
}
\startdata
1 & 10.06 & 10.01--11.00 & 1.00 & 6.99E+02 & 185.277--1.037E+04 & 1.019E+04 & 4.69E-08 & 8.90E-09 \\
2 & 11.36 & 11.01--12.00 & 1.00 & 1.46E+04 & ...--2.267E+04 & 1.230E+04 & 1.39E-10 & 2.38E-11 \\
3 & 12.42 & 12.01--13.00 & 1.00 & 2.86E+04 & ...--3.723E+04 & 1.455E+04 & 3.33E-11 & 6.27E-12 \\
4 & 13.45 & 13.01--14.00 & 1.00 & 4.46E+04 & ...--5.421E+04 & 1.699E+04 & 1.24E-11 & 2.66E-12 \\
5 & 14.43 & 14.01--15.00 & 1.00 & 6.25E+04 & ...--7.379E+04 & 1.958E+04 & 6.46E-12 & 1.44E-12 \\
6 & 15.45 & 15.01--16.00 & 1.00 & 8.36E+04 & ...--9.611E+04 & 2.232E+04 & 3.66E-12 & 9.28E-13 \\
7 & 16.46 & 16.01--17.00 & 1.00 & 1.07E+05 & ...--1.210E+05 & 2.493E+04 & 2.25E-12 & 5.64E-13 \\
8 & 17.48 & 17.01--18.00 & 1.00 & 1.35E+05 & ...--1.492E+05 & 2.816E+04 & 1.33E-12 & 3.16E-13 \\
\enddata
\end{deluxetable}

\bibliography{cr3}{}
\bibliographystyle{aasjournal}

\end{document}